\documentclass[aps,prb,twocolumn,floatfix,superscriptaddress,longbibliography]{revtex4-2}
\usepackage{color}
\usepackage{xcolor}
\usepackage{amssymb}
\usepackage{amsmath}
\usepackage{bm}
\usepackage{hyperref}
\usepackage{graphicx}
\usepackage{amsmath,amssymb,bm,epsfig,color}

\makeatother

\begin{document}
\title{Coulomb interactions in systems of generalized semi-Dirac fermions}
\author{Mohamed M. Elsayed}
\affiliation{Department of Physics, University of Vermont, Burlington, VT 05405, USA}
\author{Bruno Uchoa}
\affiliation{Department of Physics and Astronomy, Center for Quantum Research and
Technology, University of Oklahoma, Norman, OK 73019, USA}
\author{Valeri N. Kotov}
\affiliation{Department of Physics, University of Vermont, Burlington, VT 05405, USA}
\date{\today}
\begin{abstract}
Interactions have strong effects in systems with flat bands. We examine
the role of Coulomb interactions in two dimensional chiral anisotropic
quasiparticles that disperse linearly in one direction and have relatively
flat bands near the neutrality point in the other direction, dispersing
with an arbitrary positive even power law $2n\geq2$. As in the conventional
semi-Dirac case $(n=1$), we show using renormalization group that
strong logarithmic divergences in the self-energy of generalized semi-Dirac
fermions resum and lead to a restoration of linearity in the spectrum
for arbitrary $n$ over a sizable energy window in the perturbative
regime. We discuss those results in light of previous non-perturbative
large $N_{f}$ results and address the implications for physical observables. 
\end{abstract}
\maketitle

\section{Introduction}

Semi-Dirac fermions are highly anisotropic quasiparticles in two dimensions
that disperse as relativistic fermions in one direction and as massive
Galilean invariant particles in the perpendicular direction. Conventional
semi-Dirac fermions can arise at the critical point of a topological Lifshitz transition between a semimetal and an insulator, whereby there is a merger of two Dirac cones
resulting in a $2\pi$ Berry phase around the
touching point of the conduction and valence bands \cite{montambaux2009universal,Bena2011}. In the quantum
Hall regime, semi-Dirac fermions have a signature $B^{2/3}$ scaling
of the energy of the Landau levels with the applied magnetic field
\cite{dietl2008new,montambaux2009merging}, reflecting a sublinear zero-field scaling
of the density of states (DOS) with energy around the touching point.
Semi-Dirac fermions have been experimentally observed in black phosphorus
under doping \cite{kim2015observation,kim2017two}, and more recently in nodal-line semimetal ZrSiS \cite{shao2020electronic,shao2024}
and predicted to exist in strained honeycomb lattices \cite{amorim2016novel},
BEDT-TTF$_{2}$I$_{3}$ salt under pressure \cite{katayama2006pressure}, hexagonal close packed cadmium \cite{subedi2024semidiracfermionshexagonalclosepacked}, and
VO$_{2}$/TO$_{2}$ heterostructures \cite{pardo2009half,huang2015emergence}.

Short range interactions were theoretically found to drive the system
to possible modulated ordered phases in the charge, spin and superconducting
sectors, in the vicinity of a quantum critical point \cite{uchoa2017superconducting,uryszek2019quantum,uryszek2020fermionic,roy2018quantum}.
On the other hand, long range Coulomb interactions are known to strongly
renormalize the spectrum of semi-Dirac fermions. Unlike in the case
of conventional Dirac fermions in 2D \cite{kotov2012electron}, where self-energy
corrections have a single leading logarithmic divergence in the infrared,
the self-energy corrections for semi-Dirac fermions have an unconventional
leading log squared infrared divergence \cite{isobe2016emergent,cho2016novel,kotov2021coulomb}. Previous
non-perturbative calculations in the large $N_{f}$ limit, with $N_{f}$
the fermionic degeneracy, have found a strong coupling non-Fermi liquid
regime over a wide energy range followed by a perturbative marginal
Fermi liquid in the vicinity of the fixed point \cite{isobe2016emergent}. A perturbative
renormalization group calculation by some of us \cite{kotov2021coulomb} identified
an unusual regime where resummation of the log square divergences
leads to restoration of linearity to the spectrum over a wide
energy window, making semi-Dirac fermions effectively behave as conventional
anisotropic Dirac particles away from the neutrality point. 

\begin{figure}[b]
\begin{centering}
\includegraphics[scale=0.4]{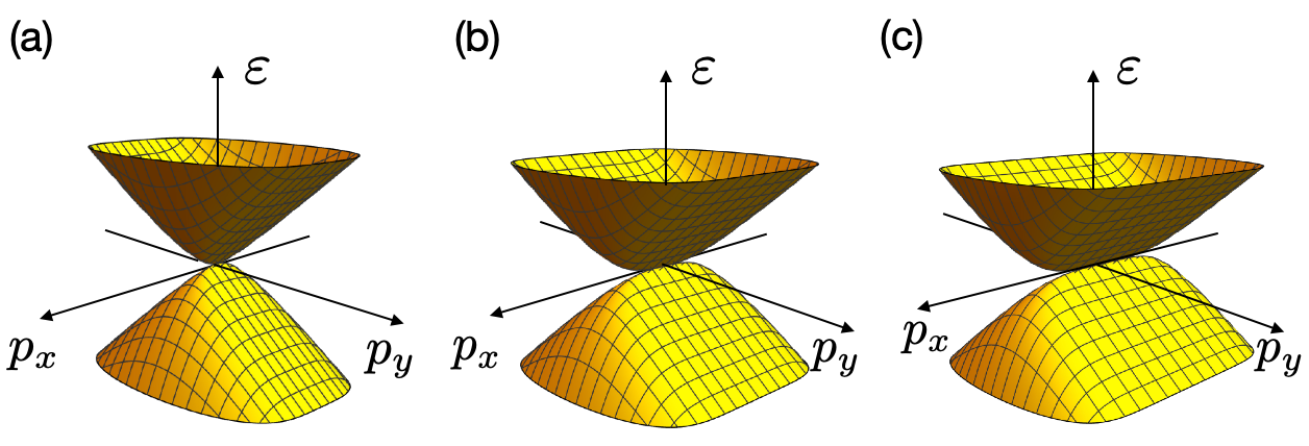}
\par\end{centering}
\caption{{\small a) Spectrum of semi-Dirac fermions ($n=1$), which disperse
linearly in one direction and parabolically in another. Generalized
semi-Dirac fermions with b) $n=2$ and c) $n=3$. The energy spectrum
becomes increasingly flat} {\small along the $p_{x}$ direction around
the touching point of the bands for larger $n$ values. The DOS scales
with energy as $\rho(\varepsilon)\propto\varepsilon^{\frac{1}{2n}}$.
(see text).}}
\end{figure}

This non-trivial resummation in the perturbative regime is very unusual, and to the best of our knowledge, has not been encountered before in other physical systems. In this paper we address two central questions: {\it i)} is this perturbative resummation unique to semi-Dirac fermions?, and {\it ii)} what is the regime of validity of the linear spectrum restoration when the number of fermionic flavors $N_f$ is of order 1? In answering {\it ii)}, we establish the crossover between the previously unconnected low energy regimes of linear renormalization and weak coupling large $N_f$ results. We show that the restoration of linearity in the perturbative regime of Coulomb interactions is not unique to semi-Dirac fermions, but is in fact present in a whole family of anisotropic 2D Hamiltonians with quasiparticles that disperse relativistically in one direction and have relatively flat bands with arbitrary curvature along the normal direction. This family of Hamiltonians has the form
\begin{equation}
\hat{\mathcal{H}}(\mathbf{p})=h_{x}(\mathbf{p})\sigma_{x}+h_{y}(\mathbf{p})\sigma_{y}\equiv\frac{g_{n}}{2}p_{x}^{2n}\sigma_{x}+vp_{y}\sigma_{y},\label{eq:H2}
\end{equation}
where $\sigma_{i}$ $(i=x,y)$ are the standard off-diagonal Pauli
matrices, and $n\in\mathbb{N}$ controls the flatness of the dispersion along the $p_{x}$ direction. $v$ is the velocity of the quasiparticles
dispersing along the $p_{y}$ direction, which are massless, and $g_{n}$
determines the curvature of the bands along the orthogonal
direction. $n=1$ corresponds to the standard dispersion for conventional
semi-Dirac fermions. The density of states for the free fermions scales as $\rho(\varepsilon)\propto\varepsilon^{\frac{1}{2n}}$
and is significantly enhanced near $\varepsilon\to0$ for $n>1$ as
the dispersion along the $p_{x}$ direction flattens out. In a strong transverse magnetic field, the Landau level spectrum scales as $\varepsilon\propto B^{2n/(2n+1)}$. 

Similar generalizations have been proposed in the literature, whereby in Ref. \cite{roy2018quantum} they study the effects of short range interactions in the one dimensional limit $n\to\infty$; in Ref. \cite{quan2018maximally} they consider the topological charge of the band crossing for different members of this class of Hamiltonians; and in Ref. \cite{carbotte2019optical} they compute optical and DC conductivities. We focus on renormalization of the spectrum in systems of generalized semi-Dirac fermions due to both the long range and screened Coulomb interaction, and investigate  the nature of the different perturbative regimes therein. This construction
provides an analytical tool to compare the effects of electron-electron
interactions in increasingly anisotropic Hamiltonians where the dispersion flattens out in one direction, remaining linear
in the other, and allows us to observe how screening is affected by the curvature of the bands.
 We show that under the bare Coulomb interaction, restoration of linearity in the spectrum previously
found for semi-Dirac fermions \cite{kotov2021coulomb} persists for arbitrary $n>1$ over an
energy window
\begin{equation}
\varepsilon_{n}<\varepsilon<\Lambda\text{e}^{-\sqrt{\frac{n}{2n-1}\pi/\alpha}},\label{eq:ineq}
\end{equation}
where $\Lambda$ is the ultraviolet energy cut-off, and $\alpha=e^{2}/v<\pi$
is a dimensionless coupling constant in the perturbative regime. The
infrared cut-off $\varepsilon_{n}\sim vq_{n}(\alpha N_f)^{2n/(2n-1)}$
signals the onset of a regime where screening effects become dominant, where $q_{n}=(2v/g_{n})^{1/(2n-1)}$ is
the characteristic momentum. In the ultra-low energy regime $\varepsilon\ll\varepsilon_{n}$
our results are in agreement with the marginal-Fermi liquid
behavior found in Ref. \cite{isobe2016emergent} for $n=1$ in the vicinity of
the weak coupling fixed point. The main results of the paper are summarized in Fig. 5. A more detailed calculation of the renormalized physical observables is given in Section \ref{Observables}.

The paper is organized in the following way: in Section \ref{Sigmas} we describe
the perturbative one-loop self energy for generalized semi-Dirac fermions.
In Section \ref{RGBare} we address the perturbative renormalization group (RG)
analysis for arbitrary $n$, where leading log square divergences
resum, leading to restoration of the linearity in the spectrum. In
Section \ref{RPA}, we examine the ultra-low energy regime of generalized
semi-Dirac fermions and in Section \ref{Observables} we calculate the renormalization
of physical observables in the zero field limit. Finally in Section \ref{Conclusions}  we present a discussion of the different perturbative regimes in
comparison with previous findings in the large $N_{f}$ limit and
present our conclusions. 

\section{Perturbative one-loop self-energy}
\label{Sigmas}

\begin{figure}[t]
\begin{centering}
\includegraphics[scale=0.55]{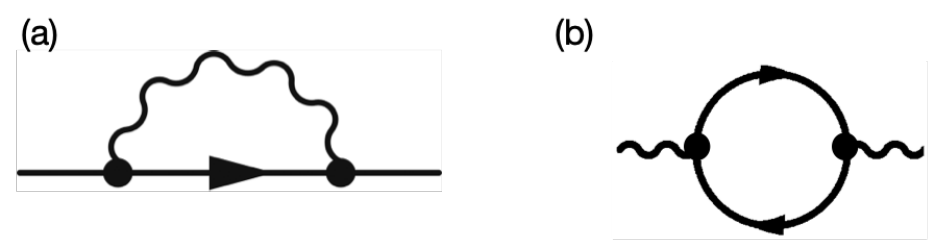}
\par\end{centering}
\caption{{\small a) Leading self-energy correction in perturbation theory. b)
One loop polarization bubble. Wavy lines are undressed Coulomb interactions.
Solid straight lines are fermionic propagators. }}
\label{Feynman}
\end{figure}

We start by calculating the one loop self-energy 
\begin{equation}
\hat{\Sigma}({\bf p})=\frac{i}{(2\pi)^{3}}\int_{-\infty}^{\infty}\text{d}\omega\int\text{d}^{2}{\bf k}\,\hat{G}_{0}({\bf k},\omega)V({\bf k}-{\bf p}),
\end{equation}
shown in Fig. \ref{Feynman}a), where 
\begin{equation}
\hat{G}_{0}({\bf k},\omega)=\left[\omega-\hat{\mathcal{H}}({\bf k})+i0^{+}\text{sgn}(\omega)\right]^{-1}
\end{equation}
is the time-ordered Green's function defined in terms of the Hamiltonian (\ref{eq:H2})
and 
\begin{equation}
V(\mathbf{k})=\frac{2\pi e^{2}}{\vert\mathbf{k}\vert}
\end{equation}
is the electron-electron Coulomb interaction. The frequency integral
is easily performed to yield: 
\begin{equation}
\hat{\Sigma}(\mathbf{p})=\Sigma_{x}(\mathbf{p})\sigma_{x}+\Sigma_{y}(\mathbf{p})\sigma_{y},\label{eq:sigma=000020matrix}
\end{equation}
where 
\begin{equation}
\Sigma_{i}(\mathbf{p})=\frac{\pi}{(2\pi)^{3}}\int\text{d}^{2}{\bf k}\frac{h_{i}(\mathbf{k})}{\sqrt{h_{x}^{2}(\mathbf{k})+h_{y}^{2}(\mathbf{k})}}\;V({\bf k}-{\bf p}),\label{eq:Sigma}
\end{equation}
with $i=x,y$. 

\subsection{Velocity self-energy correction }

We expand the potential for small external $\mathbf{p}$ ($p\ll k$), retaining terms up
to first order (the leading behavior in $p_y$) 
\begin{equation}
V({\bf k}-{\bf p})\approx\frac{2\pi e^{2}}{k}\left(1+\frac{{\bf k}\cdot{\bf p}}{k^{2}}\right).
\label{eq:vyexpansion}
\end{equation}
By parity, only the term proportional to $k_{y}p_{y}$ contributes
to the integral that gives $\Sigma_{y}(\mathbf{p})$ in Eq.(\ref{eq:Sigma}).
We transform to new coordinates defined by 
\begin{equation}
\frac{g_{n}}{2}k_{x}^{2n}=\varepsilon\sin\phi,\quad vk_{y}=\varepsilon\cos\phi,\label{transformation}
\end{equation}
such that
\begin{equation}
\varepsilon=\sqrt{\left(\frac{g_{n}}{2}k_{x}^{2n}\right)^{2}+(vk_{y})^{2}}\label{eq:epsilon}
\end{equation}
 corresponds to the energy of the quasiparticles and $0\leq\phi\leq\pi$,
since $k_{x}^{2n}\geq0$. Computing the Jacobian determinant, the
element of area is: 
\begin{equation}
\text{d}^{2}\mathbf{k}=\frac{2^{\frac{1}{2n}-1}}{nvg_{n}^{\frac{1}{2n}}}\varepsilon^{\frac{1}{2n}}(\sin\phi)^{\frac{1}{2n}-1}\,\text{d}\varepsilon\text{d}\phi.
\end{equation}
Every two points in $k$-space are mapped to a single point in the
$(\varepsilon,\phi)$ space as defined, and so one must add an additional
factor of $2$ to fully integrate over $k$-space in the new variables.
It is useful to introduce a dimensionless energy variable 
\begin{equation}
E=\left(\frac{g_{n}}{2\varepsilon}\right)^{\frac{1}{n}}\frac{\varepsilon^{2}}{v^{2}}=\left(\frac{\varepsilon}{q_{n}v}\right)^{2-\frac{1}{n}},\label{dimE}
\end{equation}
where $q_{n}=\left(2v/g_{n}\right)^{\frac{1}{2n-1}}$ has units of
momentum. In the new variables, the integral becomes: 
\begin{equation}
\Sigma_{y}(\mathbf{p})=vp_{y}\frac{\alpha}{4\pi}\int\text{d}E\,\int_{0}^{\pi}\text{d}\phi\,f(E,\phi),\label{eq:sigmay}
\end{equation}
where $\alpha=e^{2}/v$ is the dimensionless coupling constant and
\begin{equation}
f(E,\phi)=\frac{\cos^{2}\phi}{(\sin\phi)^{1-(1/2n)}\left(\sin^{1/n}\phi+E\cos^{2}\phi\right)^{3/2}}.\label{eq:f}
\end{equation}
The integral sharply accumulates around $\phi=0$ so we estimate: 
\begin{eqnarray}
\int_{0}^{\pi}d\phi\,f(E,\phi) & \approx & 2\left(\int_{0}^{a}\frac{\text{d}\phi}{\phi^{1-(1/2n)}\left[\phi^{1/n}+E\right]^{3/2}}\right.\nonumber \\
 & + & \left.\int_{a}^{\pi/2}d\phi\,f(E,\phi)\right)\label{eq:approx}\\
 & \approx & 2\int_{0}^{a}\frac{\text{d}\phi}{\phi^{1-(1/2n)}\left[\phi^{1/n}+E\right]^{3/2}}\nonumber \\
 & = & \frac{4na^{1/2n}}{E\sqrt{a^{1/n}+E}},
\end{eqnarray}
for some $a$. We neglect the second term in Eq. (\ref{eq:approx})
since it is finite as $E\to0$. Taking $E\ll a^{1/n}\ll1$ gives 
\begin{equation}
\int_{0}^{\pi}\text{d}\phi\,f(E,\phi)\underset{E\to0}{\approx}\frac{4n}{E},
\end{equation}
and thus
\begin{equation}
\Sigma_{y}(\mathbf{p})=vp_{y}\frac{\alpha}{\pi}\frac{n}{2n-1}\int_{E_{\omega_{\mathbf{p}}}}^{E_{\Lambda}}\text{d}E\,\frac{1}{E}=vp_{y}\frac{\alpha}{\pi}\ln\left(\frac{\Lambda}{\omega_{\mathbf{p}}}\right).
\end{equation}
As usual, we introduce a high energy cutoff $\Lambda$ and integrate
above an on-shell infrared energy cut-off $\omega_{\mathbf{p}}\equiv\varepsilon$,
i.e. $E_{\omega_{\mathbf{p}}}<E<E_{\Lambda}$. The effective perturbative
parameter that controls the expansion is $\bar{\alpha}\equiv\alpha/\pi<1$,
with $\alpha=e^{2}/v$ the dimensionless fine structure constant of
the material. 

\begin{figure*}
\begin{centering}
\includegraphics[scale=0.5]{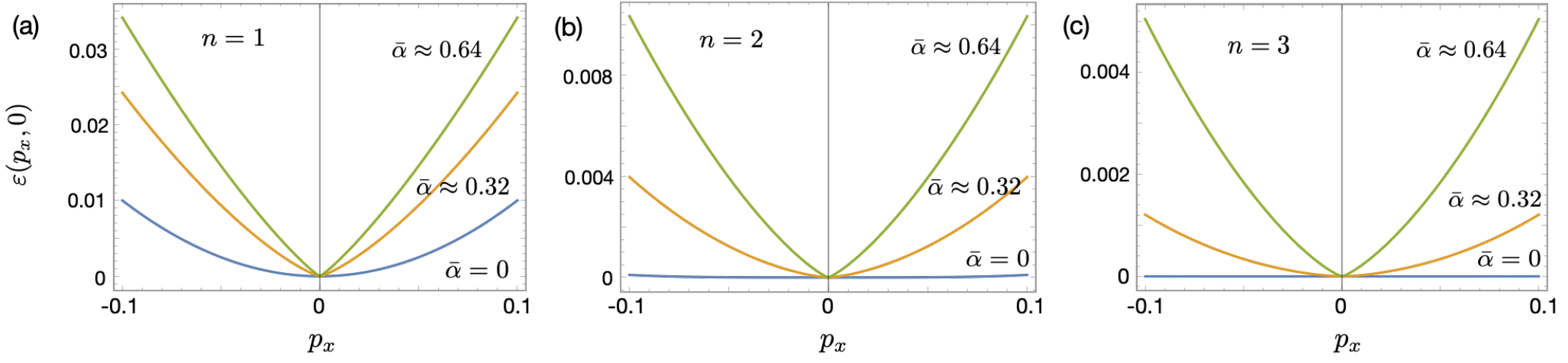}
\par\end{centering}
\caption{{\small Renormalized energy spectrum along the $p_{x}$ direction for
a) $n=1$, b) $n=2$ and $n=3$. Energy $\varepsilon$ and momentum
$p_{x}$ normalized by the ultraviolet cut-offs $\Lambda$ and $p_{\Lambda}^{x}=(2\Lambda/g_{0})^{1/2n}$
respectively. Blue lines describe the non-interacting spectrum of
generalized semi-Dirac fermions, with $\propto p_{x}^{2n}$ dispersion.
Orange lines correspond to the perturbative parameter $\bar{\alpha}\equiv\alpha_0/\pi\approx0.32$
and green lines to $\bar{\alpha}\approx0.64$. Interactions restore
the linearity of the energy spectrum along the $p_{x}$ direction
in momentum for any integer $n\protect\geq1$. The effective velocity
of the quasiparticles $v_{x}$ in the renormalized spectrum has a
weak dependence on $n$ (see text). }}
\label{LinearPlots}
\end{figure*}

\subsection{Self-energy correction to $g_{n}$}

To find the correction to $g_{n}$, we expand 
\begin{equation}
V({\bf k}-{\bf p})=\frac{2\pi e^{2}}{\vert{\bf k}-{\bf p}\vert}=2\pi e^{2}\sum_{l=0}^{\infty}\frac{p^{l}}{k^{l+1}}P_{l}(\cos\gamma),
\end{equation}
where $P_{l}$ are the Legendre polynomials and $\cos\gamma=\frac{{\bf k}\cdot{\bf p}}{kp}$.
The expansion is only valid for $k>p$, and only $l=even$ terms contribute
to the integral in Eq. (\ref{eq:Sigma}). We consider the term proportional
to $p_{x}^{2n}$ ($l=2n$), and recast into the energy-angle coordinates
in Eqs.(\ref{transformation}),(\ref{dimE}) to arrive at
\begin{equation}
\Sigma_{x}(\mathbf{p})=\frac{g_{n}}{2}p_{x}^{2n}\frac{\alpha}{4\pi(2n-1)}\int\frac{\text{d}E}{E}\int\text{d}\phi\,L(E,\phi),\label{eq:Sigma-1}
\end{equation}
where
\begin{equation}
L(E,\phi)=\frac{\sin^{\frac{1}{2n}}\phi}{\left(\sin^{1/n}\phi+E\cos^{2}\phi\right)^{n+\frac{1}{2}}}P_{2n}(\cos{\gamma}),\label{eq:L}
\end{equation}
with
\begin{equation}
\cos\gamma=\frac{\sin^{1/2n}\phi(\cos\beta)+\sqrt{E}\cos\phi(\sin\beta)}{\sqrt{\sin^{1/n}\phi+E\cos^{2}\phi}}.
\end{equation}
and $\beta$ is the polar angle of $\mathbf{p}$.
Since the leading divergence occurs as $E\to0$, we restrict the integral
to 
\begin{equation}
E\ll\frac{\sin^{1/n}\phi}{\cos^{2}\phi},
\end{equation}
allowing us to approximate $\cos\gamma\approx \cos\beta=1$, where we take $\mathbf{p}$ to be along the $x$ direction, and $L(E,\phi)\approx1/\sin\phi.$
The singularity is captured at small $\phi$ so we consider $E^{n}\ll\phi\ll1,$
where 
\begin{equation}
\int_{0}^{\pi/2}\text{d}\phi \, L(E,\phi)\approx\int_{E^{n}}\text{d}\phi\frac{1}{\phi},
\end{equation}
and ignore the finite behavior on the upper bound. Neglecting subleading
terms, we are left with 
\begin{equation}
\int_{E_{\omega}}^{E_{\Lambda}}\text{d}E\frac{-2n\ln E}{E}=n\ln^{2}\left(\frac{E_{\Lambda}}{E_{\omega}}\right),\label{eq:Int}
\end{equation}
finally yielding: 
\begin{equation}
\Sigma_{x}(\mathbf{p})=\frac{g_{n}}{2}p_{x}^{2n}\frac{\alpha}{4\pi}\frac{2n-1}{n}\ln^{2}\left(\frac{\Lambda}{\omega_{\mathbf{p}}}\right).
\end{equation}
This unconventional log squared singularity originally reported in
semi-Dirac fermions $(n=1)$ appears for all $n$. This has a significant
effect on the forthcoming RG analysis of this problem.

\section{Renormalization Group}

\label{RGBare} We set up the renormalization group (RG) equations
\begin{eqnarray}
\frac{dg}{d\ell} & = & \frac{1}{2\pi}\frac{2n-1}{n}\alpha(\ell)g(\ell)\ell\label{gRGeqn}\\
\frac{dv}{d\ell} & = & \frac{1}{\pi}\alpha(\ell)v(\ell),\label{vRGeqn}
\end{eqnarray}
where we introduce the RG scale $\ell=\ln\frac{\Lambda}{\omega_{\mathbf{p}}}$.
Since the charge does not renormalize, the
solution to Eq.(\ref{vRGeqn}) is trivial, decoupling the equations.
There are no higher order logs to sum for $v$, and $\alpha$ only
renormalizes via its dependence on $v$. The solution to Eq.(\ref{gRGeqn})
reads
\begin{equation}
\frac{g(\omega)}{g_{0}}=\left[\frac{\Lambda/\omega}{\left(1+\frac{\alpha_{0}}{\pi}\ln\frac{\Lambda}{\omega}\right)^{\pi/\alpha_{0}}}\right]^{\frac{2n-1}{2n}},
\end{equation}
with $g_{0}$ and $\alpha_{0}$ as the bare constants. The strong
power law dependence on $\omega$ has a profound effect on the spectrum.
To see this more clearly we consider the renormalized dispersion along
the two directions 
\begin{eqnarray}
\varepsilon(p_{x},0) & = & \frac{\Lambda}{p_{\Lambda}^{x}}\frac{1}{\left[1+\frac{\alpha_{0}}{\pi}\,2n\ln\left(\frac{p_{\Lambda}^{x}}{|p_{x}|}\right)\right]^{\frac{\pi}{\alpha_{0}}\frac{2n-1}{2n}}}\,\vert p_{x}\vert\label{ERGx}\\
\varepsilon(0,p_{y}) & = & v_{0}\left[1+\frac{\alpha_{0}}{\pi}\ln\left(\frac{\Lambda}{v_{0}|p_{y}|}\right)\right]\vert p_{y}\vert,\label{ERGy}
\end{eqnarray}
where we define the characteristic high-energy momentum scale
\begin{equation}
p_{\Lambda}^{x}=\left(\frac{2\Lambda}{g_{0}}\right)^{1/2n}.
\end{equation}

Remarkably, the resummation of log squared divergences to all orders of perturbation theory has produced a linear spectrum with a weak
additional logarithmic correction. Note that this behavior is general
for any $n\geq 1$, implying that interactions always drive the system
towards linearity regardless of how flat the bare spectrum is. The dependence on $n$ is quite weak, further exhibiting
the generality of our result. Our results agree with those in Ref.
\cite{kotov2021coulomb} for the case of ordinary ($n=1$) semi-Dirac fermions. 

In Fig. \ref{LinearPlots} we plot the renormalization of the energy spectrum along
the $p_{x}$ direction for $\alpha_0/\pi\approx0,\,0.32$ and $0.64$.
Panel a) shows the renormalized energy spectrum (orange and green lines)
for $n=1$, in the conventional case of semi-Dirac fermions. As found
before, the parabolic dispersion near the touching point becomes linear
(up to weak logarithmic corrections to scaling) over a sizable energy
range. This effect is even more pronounced for anisotropic bands that
are flatter along the $p_{x}$ direction in the vicinity of the neutrality
point, as shown in panels b) and c), for $n=2$ and $3$ respectively.
The effective renormalized velocity of the quasiparticles along that direction
$v_{x}=\partial_{p_x}\varepsilon(p_{x},0)$, to leading order, has a weak dependence
on $n$ at finite momentum $p_{x}\ll p_{\Lambda}^{x}$. In other words, $v_{x}(\ell)$
is weakly dependent on the form of the original free dispersion. This
is a non-trivial, non-perturbative effect that results from the resummation
of leading logs to all powers in perturbation theory in the RG flow,
even though the system is still within the perturbative regime $\alpha_0/\pi<1$. As we discuss below, as $\alpha_0/\pi\to0$,
the energy window where the restoration of linearity occurs becomes
exponentially small compared to the ultraviolet cut-off $\Lambda$.
For typical finite values $\alpha_0/\pi<1$, the energy window can be
fairly wide and is amenable to experimental observation. This is to be contrasted with resummations typically encountered in non-perturbative $1/N_f$ expansions, which result in finite corrections to the dynamical critical exponents that scale to leading order with $1/N_{f}$ \cite{isobe2016emergent,foster2008graphene,nandkishore2010electron,dou2014quasiparticle}. For instance, the large $N_f$ expansion for $n=1$ yields: 

\begin{equation}
   \frac{g(\omega)}{g_0}=\left(\frac{\Lambda}{\omega}\right)^{0.1261/N_f},\, \frac{v(\omega)}{v_0}=\left(\frac{\Lambda}{\omega}\right)^{0.3625/N_f}. 
\end{equation}

\begin{figure*}
\begin{centering}
\includegraphics[scale=0.5]{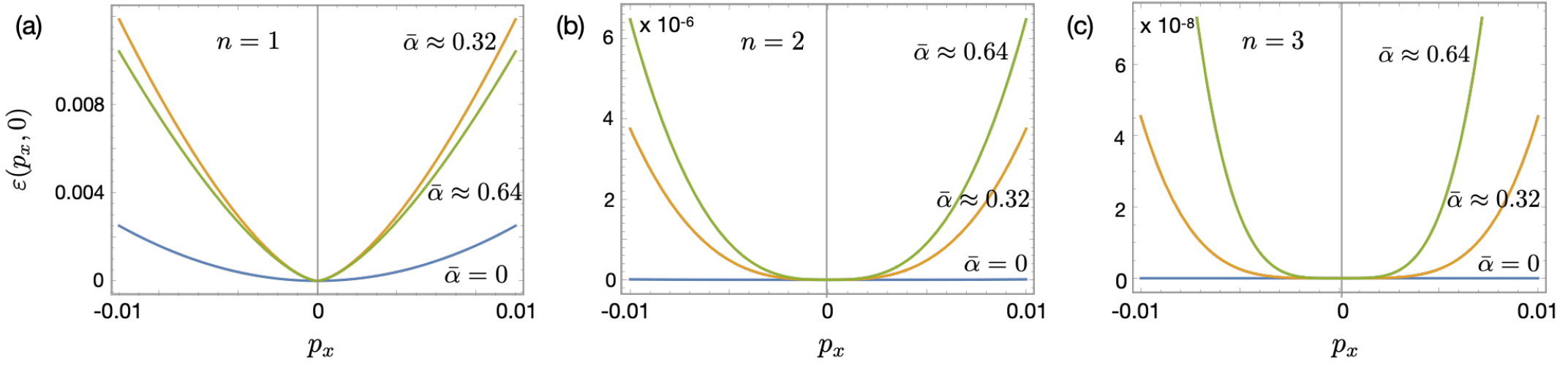}
\par\end{centering}
\caption{{\small Renormalized energy spectrum along the $p_{x}$ direction in
the RPA dominated, ultra-low energy regime ($N_f=1$) for a) $n=1$, b) $n=2$,
and c) $n=3$. Energy $\varepsilon$ and momentum $p_{x}$ normalized
by the ultraviolet cut-offs $\Lambda$ and $p_{\Lambda}^{x}=(2\Lambda/g_{0})^{1/2n}$
respectively. Blue lines: $\bar{\alpha}=0$; orange lines: $\bar{\alpha}\equiv\alpha/\pi\approx0.32$;
green lines: $\bar{\alpha}\approx0.64$. In this regime interactions
produce weaker corrections to the energy spectrum. The peculiar non-monotonic dependence of $g_n$ on $\alpha$ is clearly visible in panel a) (see text). }}
\label{RPAPlots}
\end{figure*}

\section{Ultra-low energy regime}
\label{RPA}

Because the bare density of states vanishes near the neutrality point
with sublinear power law dependence in momentum along the $k_{y}$
direction, the Coulomb interaction diverges faster with momentum in
the infrared. This requires us to consider the effects of long-distance
screening on the self-energy. The static polarization function $\Pi(\mathbf{k},\nu=0)$
along the two axes (diagrammatically shown in Fig. \ref{Feynman}b) can be calculated
analytically to give: 
\begin{align}
\Pi(k_{x},0) & =-\frac{\vert k_{x}\vert}{4\pi^{2}v}a_{n}\label{eq:Pix}\\
\Pi(0,k_{y}) & =-\frac{(\frac{2v}{g_{n}})^{\frac{1}{2n}}\vert k_{y}\vert^{\frac{1}{2n}}}{2\pi v}c_{n},\label{eq:Piy}
\end{align}
where 
\begin{equation}
a_{n}=\int_{-\infty}^{\infty}\text{d}x\left[1-\frac{x^{2n}(x+1)^{2n}}{(x+1)^{4n}-x^{4n}}\,4n\ln\left(\frac{x+1}{x}\right)\right]\label{eq:a}
\end{equation}
and 
\begin{equation}
c_{n}=\frac{\Gamma\left(1+\frac{1}{4n}\right)}{\Gamma\left(\frac{3}{2}+\frac{1}{4n}\right)}\frac{\sqrt{\pi}\csc\left(\frac{\pi}{4n}\right)}{2^{2+1/2n}\;n}.\label{eq:c}
\end{equation}
The integral in Eq.(\ref{eq:a}) can be exactly evaluated for
$n=1$; $a_{1}=\pi^2/4\approx 2.47$ , whereas $c_{1}=\sqrt{\pi}\,\Gamma(\frac{5}{4})/\Gamma(\frac{7}{4})\approx0.44$.
For the other relevant cases ($n=2,3)$, $a_{2}\approx 5.36$, $a_{3}\approx 8.19$,
$c_{2}\approx0.30$ and $c_{3}\approx0.22$.

The polarization bubble has linear in momentum scaling along the $k_{x}$
direction and sublinear momentum dependence along the $k_{y}$ direction.
Calculating the self-energy corrections under the dressed Coulomb
interaction $V_{\text{RPA}}(\mathbf{k})=V(\mathbf{k})/[1-V(\mathbf{k})\Pi(\mathbf{k})]$,
as we show in more detail in Appendix \ref{AppendixA} , we replace the bare Coulomb
propagator with the screened one. The integral appearing in Eq. (\ref{eq:Int})
is replaced with 
\begin{equation}
\int_{E_{\omega}}^{E_{\Lambda}}\text{d}E\frac{\ln\left(\sqrt{E}+c_{n}\alpha N_{f}\right)}{E},\label{eq:Int2}
\end{equation}
where $N_{f}$ is the number of fermionic species. In the ultra-low
energy regime $\sqrt{E}\ll c_{n}\alpha N_{f}$, or equivalently for
\begin{equation}
\varepsilon\ll vq_{n}(c_{n}\alpha N_{f})^{\frac{2n}{2n-1}},\label{eq:Ultra-low}
\end{equation}
screening precludes the emergence of a log square term. In this regime,
the corrections are: 
\begin{eqnarray}
v & \to & v\left[1+\frac{\alpha}{2\pi n}\ln\left(\frac{\Lambda}{\omega_{\mathbf{p}}}\right)\right]\\
g_{n} & \to & g_{n}\left[1+\frac{\alpha}{\pi}\ln\left(\frac{1}{c_n\alpha N_f}\right) \ln\left(\frac{\Lambda}{\omega_{\mathbf{p}}}\right)\right]
\end{eqnarray}
The correction to $v$ has now acquired an $n$ dependent coefficient. More importantly, the RPA screening has `split' the log square, leaving
a single log divergence with a non-perturbative coefficient. Note
that this has been studied in detail for the $n=1$ case in Ref. \cite{isobe2016emergent}
, where they calculate the log coefficients to be twice what we find. Nevertheless,
it is the ratio of coefficients that determines the RG flow and hence
the forthcoming results are in exact agreement with their work near
the fixed point of the problem. 

We proceed as in Section \ref{RGBare}, and find that again the velocity
does not run in the ultra-low energy regime, whereas
\begin{equation}
\frac{g(\ell)}{g_{0}}=\left(1+\frac{\alpha_{0}}{2\pi n}\ell\right)^{\gamma_{n}(\ell)},
\end{equation}
with the exponent
\begin{equation}
\gamma_{n}(\ell)=n\ln\left[\frac{1+\frac{\alpha_{0}}{2\pi n}\ell}{(c_{n}\alpha_0 N_{f})^{2}}\right].\label{eq:gamma}
\end{equation}
The correction to $g$ has lost the power law in $\omega$, but has
acquired a particularly strong logarithmic correction. The effect
is much more sensitive to the value of $n$, and less sensitive to
the value of $\alpha_0$. There is no general behavior for all $n$,
but the the spectrum is still significantly enhanced:
\begin{eqnarray}
\varepsilon(p_{x},0) & = & \frac{g_0}{2} \left(1+\frac{\alpha_{0}}{2\pi n}\ell\right)^{\gamma_{n}(\ell)} p_{x}^{2n}\label{ERGSCx}\\
\varepsilon(0,p_{y}) & = & v_{0}\left[1+\frac{\alpha_{0}}{2\pi n}\ln\left(\frac{\Lambda}{v_{0}\vert p_{y}\vert}\right)\right]\vert p_{y}\vert.\label{ERGSCy}
\end{eqnarray}
The effect is suppressed in the $y$ direction with increasing $n$,
but in the $x$ direction the behavior is more subtle, and is determined
by the interplay between $\alpha_{0}$, $n$, and $N_f$. We show the renormalized energy spectrum along the $p_{x}$ direction
for the ultra-low energy regime in the three panels of Fig. \ref{RPAPlots} for
$n=1,\,2$ and $3$. For fixed $n,N_f,$ and $p_x$, the spectrum is enhanced up to a critical value of
$\alpha_{0}$ (approximately $\alpha_{0}\approx n/N_f$), then begins to saturate as
$\alpha_{0}$ increases further. This can be seen clearly in Fig.
\ref{RPAPlots}a), for $n=1, N_f=1$ where the maximally enhanced spectrum occurs at $\bar{\alpha}_{0}\equiv\alpha_{0}/\pi\approx0.3$.

As the RG flows towards the non-interacting fixed point of the theory
at $\alpha\to0$, the linearity of the energy spectrum is restored
for 
\begin{equation}
\varepsilon\lesssim\Lambda\text{e}^{-\sqrt{\frac{n}{2n-1}\pi/\alpha}},\label{eq:upperbound}
\end{equation}
setting an upper bound on energy past which $\frac{\alpha}{\pi}\frac{2n-1}{n}\ln^{2}\left(\frac{\Lambda}{\omega}\right) \lesssim 1$, and the RG resummation is not justified. Therefore the region $\varepsilon>\Lambda\text{e}^{-\sqrt{\frac{n}{2n-1}\pi/\alpha}}$
corresponds to the high-energy free fermion regime, where the spectrum is unrenormalized. The restoration
of linearity of the spectrum persists all the way down to a second
energy scale, where screening effects become dominant.\textbf{ }This
low energy scale is set by Eq. (\ref{eq:Ultra-low}), 
\begin{equation}
\varepsilon_{n}=vq_{n}(c_{n}\alpha N_{f})^{\frac{2n}{2n-1}}.\label{eq:epsilonn}
\end{equation}
Below $\varepsilon\lesssim\varepsilon_n$,the system is in the ultra-low energy RPA dominated regime.

\begin{figure}[t]
\begin{centering}
\includegraphics[scale=0.45]{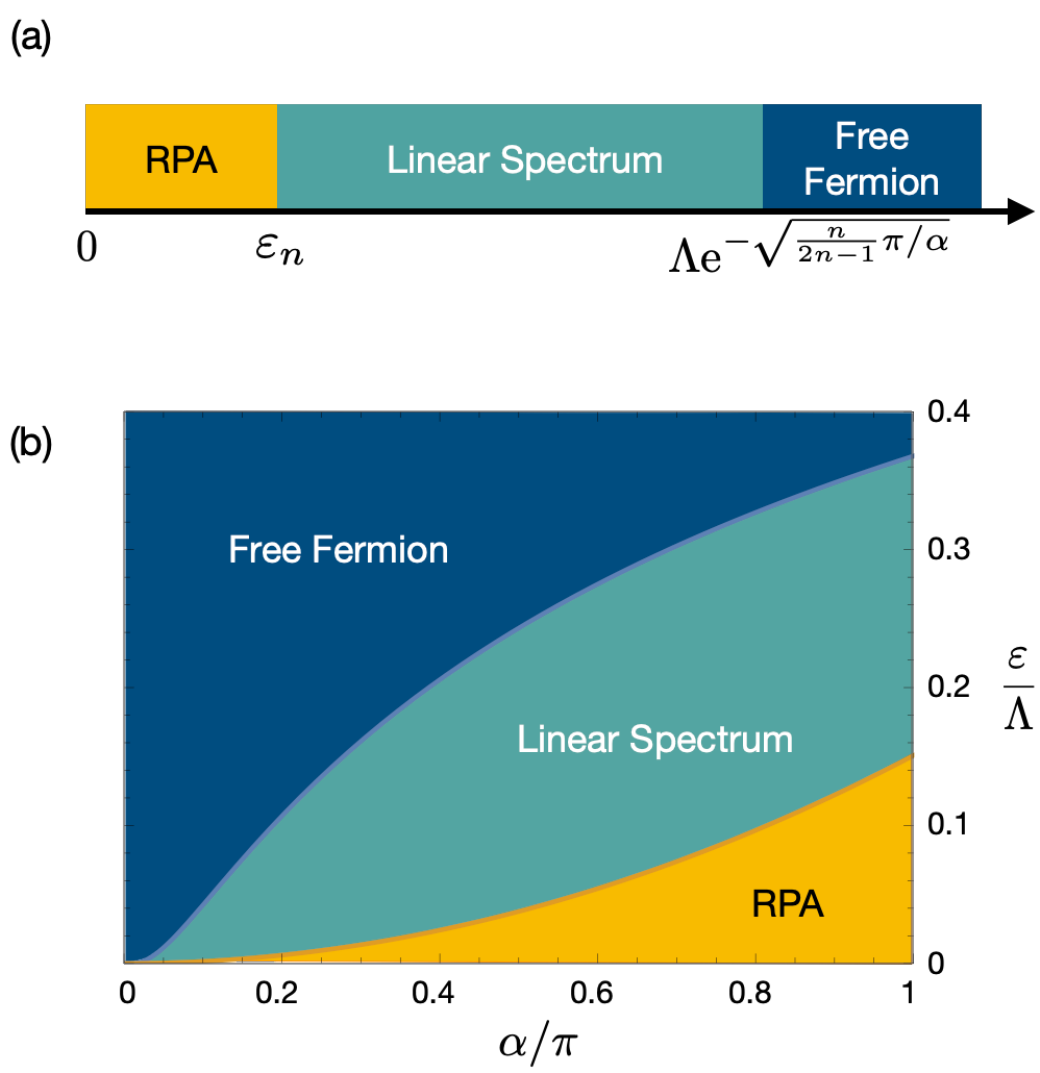}
\par\end{centering}
\caption{{\small Different weak coupling regimes in the RG flow for generalized
semi-Dirac fermions. a) The electrons are unrenormalized (free fermion)
at high energies (dark blue region). In the intermediate regime (light
green) $\varepsilon_{n}\lesssim\varepsilon\lesssim\Lambda\text{e}^{-\sqrt{\frac{n}{n-1}\pi/\alpha}}$,
resummation of log square divergences leads to the restoration of
the linearity in the spectrum, regardless of the strength of the coupling
$\alpha/\pi\lesssim1$ or the flatness of the bands (indicated by
$n$). The yellow area indicates the ultra-low energy regime $\varepsilon\ll\varepsilon_{n}\sim vq_{n}(\alpha N_f)^{2n/(2n-1)}$
(see text) , where RPA screening is dominant. In this regime, interactions
produce a particularly strong logarithmic renormalization of the spectrum, but do not reduce the power law. b) Plot
of the boundaries between different regimes versus the perturbative
coupling $\alpha/\pi$ for $n=1$ and $N_{f}=4$. The boundaries are
weakly dependent on $n$ and have a qualitatively similar shape for
$n=2$ and 3. The energy window of the regime with linear spectrum
is suppressed in the non-interacting limit $\alpha\to0$, but extends
over a wide energy range for finite coupling $\alpha/\pi<1$. The
lower bound of the window increases with the number of fermionic species
$N_{f}$ as $\propto N_{f}^{2n/(2n-1)}$. The RPA dominated regime
takes over the intermediate linear spectrum regime in the large $N_{f}$
limit, but remains subdominant to the linear spectrum regime when
$N_{f}$ is of order $1$ over a wide range of energies. }}
\label{Regimes}
\end{figure}

This behavior is illustrated in Fig. \ref{Regimes}, where the different weak coupling
regimes of the RG flow are shown. The dark blue area corresponds to
the high-energy, free fermion regime, followed by an intermediate
regime (light green) with restoration of linearity in the spectrum.
The deep infrared RPA dominated regime is shown in yellow. In this regime, 
screened Coulomb interactions produce a logarithmic renormalization
to the energy spectrum. 

In panel \ref{Regimes}b) we quantitatively show the boundaries between the different
regimes for $n=1$ and $N_{f}=4$ as a function of the perturbative
coupling $\alpha/\pi<1$. The boundaries are weakly dependent on $n$
and retain the same qualitative features for $n=2,3$ as in the $n=1$
case. For $N$ of order $1$, the energy window of the linear spectrum
regime is wide at finite coupling $\alpha/\pi$ and persists all the
way down to the weak coupling $\alpha\to0$ limit, although it becomes
exponentially small. The lower bound of the
linear regime, $\varepsilon_{n}$, increases with the fermionic degeneracy
$N_{f}$ as $\varepsilon_{n}\propto N_{f}{}^{\frac{2n}{2n-1}}$, raising
the slope of the curve that delimits the RPA region from above (yellow
region). In the large $N_{f}$ limit, considered in Ref. \cite{isobe2016emergent}
for the $n=1$ case, the RPA region takes over the linear spectrum
regime, and a non-perturbative analysis of the problem is required.
For finite $N_{f}$, and in specific for $N_{f}$ of order 1, restoration
of the linearity of the spectrum is present over a significant energy
range in units of the ultraviolet cut-off $\Lambda$. We discuss our
results in light of the existing literature in Section \ref{Conclusions}. 

\section{Physical Observables}
\label{Observables}

\subsection{Density of States}
\label{DOSSection}
We now calculate the density of states
\begin{equation}
\rho(\omega)=\frac{1}{(2\pi)^{2}}\int\text{d}^{2}\mathbf{k}\,\delta(\omega-\varepsilon(\mathbf{k}))
\end{equation}
for the renormalized spectra in the three regimes shown in Fig. \ref{Regimes}.
We approximate that the logarithmic corrections are slowly-varying
functions and thus may be considered constant over the integral. However,
we retain the strong power law correction in the case of the long-range
interaction. This yields an anisotropic Dirac cone with an effective
velocity in the $x$ direction,
\begin{equation}
v_{x}(\ell)=\frac{\Lambda}{p_{\Lambda}^{x}}\left(1+\frac{\alpha_{0}}{\pi}\ell\right)^{-\frac{\pi}{\alpha_{0}}\frac{2n-1}{2n}},
\end{equation}
This velocity is affected by weak logarithmic corrections, $\ell=\ln(\Lambda/\omega)$.
The resulting density of states as a function of
energy is:
\begin{equation}
\rho(\omega)=\frac{p_{\Lambda}^{x}}{2\pi v_{0}}\frac{\omega}{\Lambda} \left[1+\frac{\alpha_{0}}{\pi}\ln\left(\frac{\Lambda}{\omega}\right)\right]^{\left(\frac{2n-1}{2n}\right)\frac{\pi}{\alpha_{0}}-1},
\end{equation}
recovering the linearity of the DOS for conventional Dirac fermions. 

In the ultra-low energy regime, where interactions are partially screened
by polarization effects, $g(\omega)$ does not have a power law dependence.
Hence we simply calculate the density of states for the free model
and restore the logarithmic corrections after the integral is evaluated.
The result is
\begin{equation}
\rho(\omega)=\frac{1}{(2\pi)^{2}v_{0}}\left(\frac{2}{g_{0}}\right)^{\frac{1}{2n}}\negmedspace\omega^{\frac{1}{2n}}\frac{d_{n}}{n}\left[1+\frac{\alpha_{0}}{2\pi n}\ln\left(\frac{\Lambda}{\omega}\right)\right]^{-\delta_{n}(\omega)}
\end{equation}
where
\begin{equation}
\delta_{n}(\omega)= 1+\frac{1}{2}\ln\left[\frac{1+\frac{\alpha_{0}}{2\pi n}\ln\left(\frac{\Lambda}{\omega}\right)}{(c_n\alpha_{0} N_f)^{2}}\right],\label{eq:delta}
\end{equation}
and
\begin{equation}
d_{n}=\frac{1}{2^{1-1/2n}\Gamma\left(\frac{1}{2n}\right)}\left(\frac{\pi\csc\left(\frac{\pi}{4n}\right)}{\Gamma\left(1-\frac{1}{4n}\right)}\right)^{2}.\label{eq:d}
\end{equation}
Note that taking the limit $\alpha_{0}/\pi\to0$ in both expressions
recovers the correct $\omega^{1/2n}$ dependence for the free density of states,
but the coefficients do not agree. The RG flow to linearity has significantly
reduced the anisotropy in the system, producing a coefficient independent
of $n$. In the screening regime, angular integrals generate the dependence
$d_{n}/n$. 

\begin{figure}[t]
\begin{centering}
\includegraphics[scale=0.43]{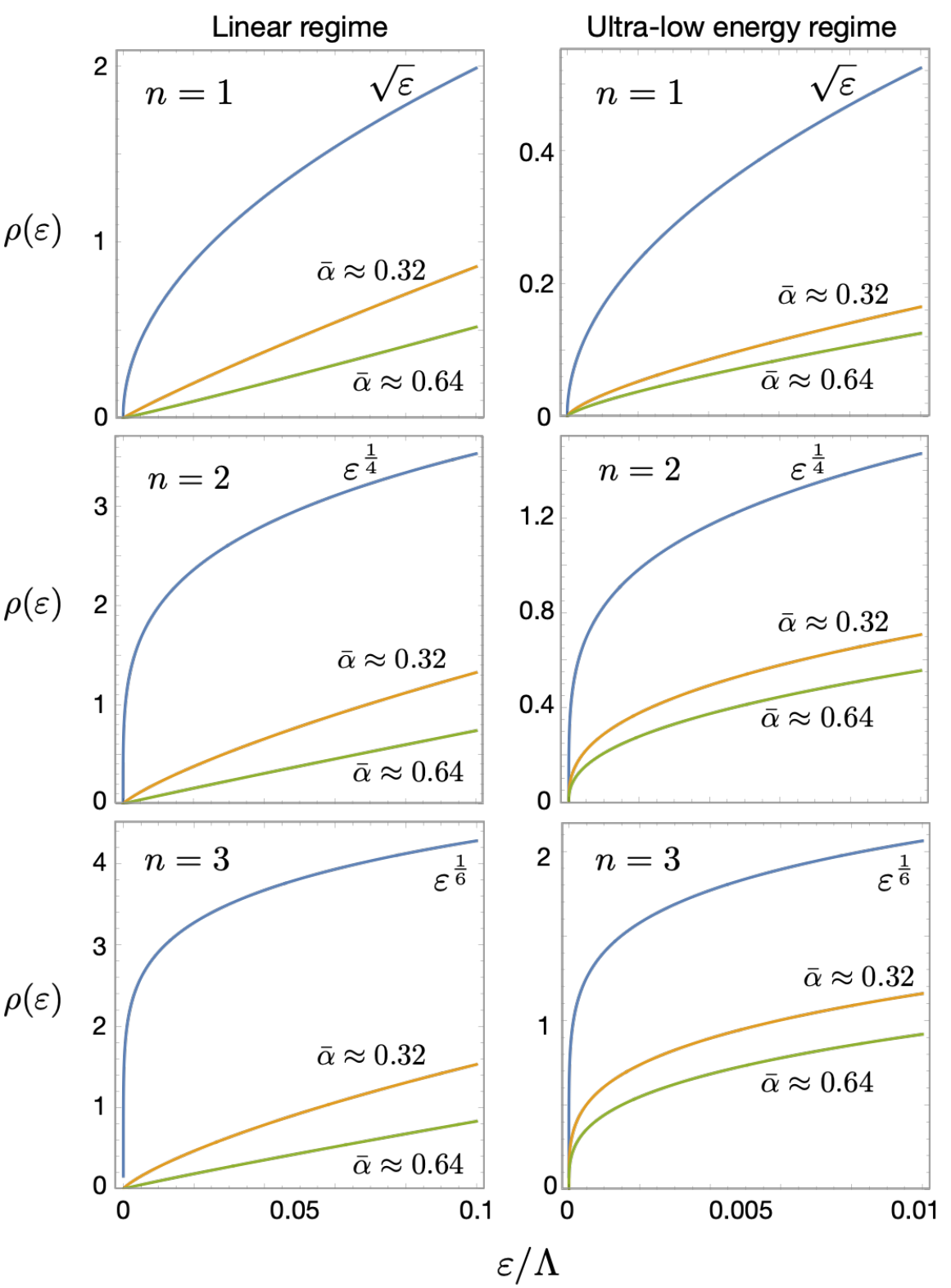}
\par\end{centering}
\caption{{\small Density of states vs energy $\varepsilon$ in the perturbative
linear regime (left column) and in the ultra-low energy regime (right
column) for $n=1$ (top panels) $n=2$ (middle) and $n=3$ (bottom).
Blue lines: bare energy spectrum for generalized semi-Dirac fermions,
with $\rho\propto\varepsilon^{1/2n}$. Orange: $\bar{\alpha}\approx0.32$;
green $\bar{\alpha}\approx0.64$. Under the bare Coulomb interaction, the crossover from $\propto\varepsilon^{1/2n}$ to a linear dependence on energy occurs quickly as interactions are turned on. In the RPA dominated ultra-low energy regime, increasing interaction strength is less impactful, and particularly so for increasing $n$. This is partly due to the $1/n$ factor incurred in $\Sigma_y$. The density of states has units of $p_{\Lambda}^{x}/2\pi v_0$ and $p_{\Lambda}^{x}/(2\pi)^2 v_0$ in the left and right columns respectively, and energy is normalized by the ultraviolet cut-off $\Lambda$.}}
\label{DOSPlots}
\end{figure}

The effect of the renormalization of the energy spectrum in the DOS
is shown in Fig. \ref{DOSPlots}. The non-interacting DOS for $n=1,2,3$ is shown
in the blue curves, where $\rho\propto\varepsilon^{\frac{1}{2n}}$.
Orange and green curves correspond to $\alpha/\pi\approx0.32$ and
$0.64$ respectively. On the left column, we show the renormalized
DOS in the intermediate linear regime, where the DOS scales linearly
with $\varepsilon$ up to logarithmic corrections to scaling. On the
right column, we show the DOS in the ultra-low energy regime, where
the DOS has modest logarithmic corrections that are consistent with
marginal Fermi liquid behavior. 

\subsection{Heat Capacity}
\label{CVSection}

The heat capacity is defined as $C_{V}=-T\partial^{2}F/\partial T^{2}$,
where $F$ is the thermodynamic free energy and we calculate it via 
\begin{equation}
C_{V}(T)=\frac{1}{(2\pi)^{2}}\frac{1}{T^{2}}\int\text{d}^{2}\mathbf{k}\left(\frac{\varepsilon(\mathbf{k})}{\cosh[\varepsilon(\mathbf{k})/2T)]}\right)^{2}.
\end{equation}
Adopting the same approach as in subsection \ref{DOSSection}, we find the heat capacity
in the unscreened case to be 
\begin{equation}
C_{V}(T)=\frac{\zeta(3)}{9\pi} \frac{p_{\Lambda}^{x}}{v_{0}\Lambda}T^{2}\left[1+\frac{\alpha_{0}}{\pi}\ln\left(\frac{\Lambda}{T}\right)\right]^{\left(\frac{2n-1}{2n}\right)\frac{\pi}{\alpha_{0}}-1}
\end{equation}
where $\zeta$ is the Riemann-zeta function. Whereas the unrenormalized
heat capacitance for generalized semi-Dirac fermions scales as $C_{V}\sim T^{1+\frac{1}{2n}}$,
restoration of  linearity in the spectrum recovers the usual
dependence for Dirac fermions, $C_{V}\sim T^{2}$, up to weak logarithmic
corrections. 

In the ultra-low energy regime, the heat capacity changes to 
\begin{equation}
C_{V}(T)=\frac{(2/g_0)^{1/2n}}{(2\pi)^{2}v_{0}} T^{1+\frac{1}{2n}}\frac{d_{n}}{n}b_{n}\left[1+\frac{\alpha_{0}}{2\pi n}\ln\left(\frac{\Lambda}{T}\right)\right]^{-\nu_{n}(T)}
\end{equation}
where 
\begin{equation}
\nu_{n}(T)=1+\frac{1}{2}\ln\left[\frac{1+\frac{\alpha_{0}}{2\pi n}\ln\left(\frac{\Lambda}{T}\right)}{(c_n \alpha_{0} N_f)^{2}}\right],\label{eqnu}
\end{equation}
and
\begin{equation}
b_{n}=\int_{0}^{\infty}\text{d}x\frac{x^{2+\frac{1}{2n}}}{\left(\cosh\frac{x}{2}\right)^{2}},
\end{equation}
where $b_{1}\approx 11.53$, $b_{2}\approx 8.63$ and $b_{3}\approx 7.87$. 

\subsection{Landau Levels}
\label{LLs}
Finally, we consider the behavior in a strong transverse magnetic field $B$ by applying the semiclassical quantization condition $S(\varepsilon)=2\pi(N+\frac{1}{2})eB$,
where $S(\varepsilon)=(2\pi)^2\int_0^{\varepsilon}\text{d}\omega\,\rho(\omega)$ is the area enclosed by a constant energy
contour in momentum space for the zero-field spectrum \cite{montambaux2009universal}. The semiclassical approximation has been checked against an exact numerical solution for the case $n=1$, and has been found to be in excellent agreement \cite{dietl2008new}. Note that the relative phase of $1/2$ cannot be obtained from semiclassical arguments, and arises as a consequence of the zero Berry phase around the semi-Dirac point. The resultant energy
spectrum of Landau levels for non-interacting generalized semi-Dirac fermions is
\begin{equation}
\label{eq:LL}
\varepsilon_N=\pm\left[v\left(\frac{g_n}{2}\right)^{\frac{1}{2n}}\frac{2n+1}{2d_{n}}2\pi\left(N+\frac{1}{2}\right)eB\right]^{\frac{2n}{2n+1}},
\end{equation}
with $N\in\mathbb{N}$. Renormalization of the Landau levels in the linear regime recovers the $\sqrt{v_x(\ell) v_y(\ell)B}$ scaling for anisotropic Dirac fermions with effective velocity $v_x(\ell)$ and $v\to v(\ell)\equiv v_y(\ell)$ as in Section \ref{RGBare}. In the ultra-low energy regime, we simply renormalize the parameters in Eq.(\ref{eq:LL}) to $g_n\to g_n(\ell), v\to v(\ell)$ as in Section \ref{RPA}.  In the presence of a magnetic field, $\omega$ in $\ell=\ln(\Lambda/\omega)$ is bounded from below by an infrared energy cutoff on the order of the spacing between Landau levels.
\section{Conclusions}
\label{Conclusions}

In summary, we examined the many-body renormalization of generalized
semi-Dirac fermions, which disperse linearly in one direction and
have nearly flat bands near the neutrality point in the perpendicular
direction with momentum dependence $\propto p_{x}^{2n}$. For larger
values of $n>1$ the spectrum becomes increasingly flat around
the touching points of the bands. We showed that the regime with restoration
of linearity of the spectrum along the $p_{x}$ direction is present
for arbitrary $n>1$, and is observable over a relatively large energy
window. This remarkable lifting of the semi-Dirac anisotropy for arbitrarily flat bands is a severe effect driven by long-range interactions.

Previous large $N_{f}$ results identified essentially three different
scaling regimes in the RG flow, as the dimensionless coupling $\alpha$
gets renormalized towards the weak coupling fixed point of the problem:
a marginal Fermi liquid regime at very low energies, which coincides
with the RPA regime we discussed for $n=1$, a high energy, free fermion
regime, and an intermediate non-Fermi liquid regime that emerges from
the large $N_{f}$ calculation. We point to the existence of a
new regime with anisotropic linear dispersion that follows from the
non-trivial resummation of leading log square divergences to all orders
in perturbation theory. This phase takes precedence over the non-Fermi
liquid regime of the large $N_f$ calculation in the situation where
$\alpha/\pi<1$ and $N_{f}$ is of order 1, when conventional perturbation
theory applies. We showed this regime appears over a wide energy
window and could be experimentally observed for instance for $n=1$.
As $N_{f}$ becomes large, the energy window progressively shrinks
as the lower bound for the linear spectrum regime grows. 

As for the realization of generalized semi-Dirac fermions with $n>1$, we expect that it might be theoretically possible in the following way. When two Dirac cones merge in a tight binding lattice model, the linear term in the low energy expansion vanishes along one direction and the leading term is quadratic in momentum, where the effective mass is a sum over hopping integrals and lattice vectors \cite{montambaux2009merging,Bena2011}. If the underlying lattice geometry and hoppings are such that this sum vanishes, then the leading behavior becomes $\propto p_x^4$, realizing the $n=2$ case in this class of Hamiltonians, and so forth. Depending on the microscopic model and the symmetry group of the lattice, this cancellation, even though  not guaranteed, can be possible in principle. In addition, it is well known that even for $n=1$ a non-universal mass (gap) term is perturbatively  generated \cite{kotov2021coulomb}; consequently  one has to theoretically ``fine tune" the system,
i.e. adjust the microscopic parameters so that it remains quantum critical. The above effects depend on the microscopic lattice realization of the class of Hamiltonians under study and are beyond the scope of the present work; most importantly the continuum low-energy analysis presented in this work and our conclusions remain valid.
Finally, we  mention that the diagram in Fig. \ref{Feynman}a)
also generates lower-power even terms, not originally present in the Hamiltonian, i.e. for the case $n=2$, a term $\propto t p_x^2 \sigma_x$ appears, with $t$ being a non-universal, cutoff dependent constant. Such terms
are not divergent and, naturally, do not affect our RG analysis  and the conclusions of the present work.

The signature features of the many-body
effects we describe can be observed in angle-resolved photoemission
experiments \cite{bostwick2007quasiparticle}, which can probe the renormalization of the quasiparticle
energy spectrum; and also in quantum capacitance measurements of the
electronic compressibility \cite{martin2008observation,yu2013interaction}, which can probe the
temperature dependence of the heat capacitance. Moreover, the signature $B^{2n/2n+1}$ scaling of the Landau levels can be detected via magneto-optical spectroscopy techniques, as presented in Ref.\cite{shao2024} for the nodal line semimetal ZrSiS.

\acknowledgments
We would like to thank A. Chubukov for enlightening discussions. BU
acknowledges NSF grant No DMR-2024864 for support. MME and VNK gratefully acknowledge the financial 
support from NASA Grant No. 80NSSC19M0143, and the CEMS Summer Graduate Fellowship at the University of Vermont. 

\appendix

\section{Screened interactions}
\label{AppendixA}

We use an ansatz for the static polarization bubble for generalized semi-Dirac fermions that recovers the exact results along the two directions:
\begin{equation}
\Pi(k_{x},k_{y})=-\left[(f_{n})^{4n}k_{x}^{4n}+\left(h_{n}\right)^{4n}k_{y}^{2}\right]^{\frac{1}{4n}},\label{eq:Bubble}
\end{equation}
where 
\begin{equation}
f_{n}=\frac{1}{4\pi^{2}v}a_{n},\qquad h_{n}=\frac{1}{2\pi v}\left(\frac{2v}{g_{n}}\right)^{\frac{1}{2n}}c_{n},\label{eq:fh}
\end{equation}
with $a_{n}$ and $c_{n}$ dimensionless constants defined in Eq.
(\ref{eq:a}) and (\ref{eq:c}). Using the parametrization (\ref{transformation}),
\begin{equation}
\frac{g_{n}}{2}k_{x}^{2n}=\varepsilon\sin\phi,\quad vk_{y}=\varepsilon\cos\phi,\label{eq:par}
\end{equation}
with $\phi\in[0,\pi]$, then $\Pi(\varepsilon,\phi)=-\left(\frac{2\varepsilon}{g_{n}}\right)^{\frac{1}{2n}}\pi_{0}(\phi$),
with 
\begin{equation}
\pi_{0}(\phi)=\left[(f_{n})^{4n}\sin^{2}\phi+\left(h_{n}^{\prime}\right)^{4n}\cos^{2}\phi\right]^{\frac{1}{4n}},\label{eq:pio}
\end{equation}
where $h_{n}^{\prime}=c_{n}/(2\pi v)$. The screened Coulomb interaction
becomes 
\begin{align}
V_{\text{RPA}}(\mathbf{k}) & =\frac{2\pi e^{2}}{\vert\mathbf{k}\vert-2\pi v\alpha N_f\Pi(\mathbf{k})}\nonumber \\
 & =\frac{2\pi e^{2}\left(\frac{g_{n}}{2\varepsilon}\right)^{\frac{1}{2n}}}{\left[\sqrt{\sin^{\frac{1}{n}}\phi+E\cos^{2}\phi}+2\pi v\alpha N_f\pi_{0}(\phi)\right]},\label{eq:Vrpa}
\end{align}
where $E=\left[\varepsilon/(q_{n}v)\right]^{2- 1/n}$ is dimensionless, and $q_{n}=\left(2v/g_{n}\right)^{\frac{1}{2n-1}}$ has units of momentum. 
Expanding $\Sigma_x$ in Eq.(\ref{eq:Sigma}) for small external $\mathbf{p}$ and transforming momentum integration variables to $(\varepsilon,\phi)$ coordinates, we find that the term
\begin{equation}
    \frac{h_{x}(\mathbf{k+p})}{\sqrt{h_{x}^{2}(\mathbf{k+p})+h_{y}^{2}(\mathbf{k+p})}}\approx \frac{g_n}{2}p_x^{2n}\frac{\cos^{4n}\phi}{\varepsilon}
\end{equation}
produces the most singular contribution to the integral.

Hence, 
\begin{align}
\Sigma_{x}(\mathbf{p}) & =\frac{g_{n}}{2}p_{x}^{2}\frac{\alpha}{4\pi}\frac{1}{2n-1}\int_{E_{\omega}}^{E_{\Lambda}}\frac{\text{d}E}{E}\int_{0}^{\pi}\text{d}\phi\frac{\cos^{4n}\phi}{(\sin\phi)^{\frac{2n-1}{2n}}}\nonumber \\
 & \qquad\times\frac{1}{\left[\sqrt{\sin^{\frac{1}{n}}\phi+E\cos^{2}\phi}+2\pi v\alpha N_f\pi_{0}(\phi)\right]}.\label{eq:sigma2}
\end{align}

The dominant contribution in the integrand occurs for small $\phi$
in the limit of $E\to0$, $(E^{n}\ll\phi)$. Approximating $\sin\phi\sim\phi$, $\cos\phi\sim 1$
the integral in $\phi$ becomes 
\begin{equation}
2\int_{E^{n}}^{\pi/2}\text{d}\phi\frac{1}{\phi^{\frac{2n-1}{2n}}}\frac{1}{\phi^{\frac{1}{2n}}+2\pi v\alpha N_f h_{n}^{\prime}}=-2n\ln\left(\sqrt{E}+\alpha^{\prime}\right).
\label{eq:intphig}
\end{equation}
up to a constant, and $\alpha^{\prime}=\alpha N_f c_{n}$. Hence, in the ultra-low energy
regime, where $\sqrt{E}\ll\alpha^{\prime}$,\textbf{ }
\begin{align}
\Sigma_{x}(\mathbf{p}) & =\frac{g_{n}}{2}p_{x}^{2}\frac{\alpha}{4\pi}\frac{-4n}{2n-1}\int_{E_{\omega}}^{E_{\Lambda}}\frac{\text{d}E}{E}\ln\left(\sqrt{E}+\alpha^{\prime}\right)\nonumber \\
 & \approx\frac{g_{n}}{2}p_{x}^{2}\frac{\alpha}{\pi}\ln\left(\frac{1}{\alpha^{\prime}}\right)\ln\left(\frac{\Lambda}{\omega}\right).\label{eq:Sigma3}
\end{align}

To find the velocity correction $\Sigma_y$ we expand $V_{\text{RPA}}(\mathbf{k-p})$ to linear order in $p_y$ similarly to Eq.(\ref{eq:vyexpansion}). The term that generated the singularity for the bare interaction is now finite as $E\to 0$, and the leading divergence is now produced by
\begin{align}
    \Sigma_y & =v p_y\frac{\alpha}{4\pi}\frac{\alpha^{\prime}}{2n(2n-1)}\int_{E_{\omega}}^{E_{\Lambda}} \frac{\text{d}E}{E}\int_0 ^{\pi}\text{d}\phi\frac{\cos^{1/2n}\phi}{(\sin\phi)^{\frac{2n-1}{2n}}} \nonumber \\
    & \qquad \times \frac{1}{\left(\sqrt{\sin^{1/n}\phi+E\cos^2\phi}+\alpha^{\prime}\cos^{1/2n}\phi\right)^2}
\end{align}
Approximating the $\phi$ integral in a similar fashion to Eq.(\ref{eq:intphig}) we arrive at
\begin{align}
    \Sigma_y & =v p_y\frac{\alpha}{4\pi}\frac{2\alpha^{\prime}}{2n-1}\int_{E_{\omega}}^{E_{\Lambda}} \frac{\text{d}E}{E}\frac{1}{\sqrt{E}+\alpha^{\prime}}\\
    & = v p_y \left(\frac{\alpha}{2\pi n}\ln\frac{\Lambda}{\omega}\right).
\end{align}
The effect of screening on the renormalization of $v$ is much milder than the splitting of the log squared in $\Sigma_x$, but it does change the coefficient numerically and generates a $1/n$ dependence.

\nocite{apsrev42Control}
\bibliographystyle{apsrev4-2}
\bibliography{references}

\begin{thebibliography}{29}%
\makeatletter
\providecommand \@ifxundefined [1]{%
 \@ifx{#1\undefined}
}%
\providecommand \@ifnum [1]{%
 \ifnum #1\expandafter \@firstoftwo
 \else \expandafter \@secondoftwo
 \fi
}%
\providecommand \@ifx [1]{%
 \ifx #1\expandafter \@firstoftwo
 \else \expandafter \@secondoftwo
 \fi
}%
\providecommand \natexlab [1]{#1}%
\providecommand \enquote  [1]{``#1''}%
\providecommand \bibnamefont  [1]{#1}%
\providecommand \bibfnamefont [1]{#1}%
\providecommand \citenamefont [1]{#1}%
\providecommand \href@noop [0]{\@secondoftwo}%
\providecommand \href [0]{\begingroup \@sanitize@url \@href}%
\providecommand \@href[1]{\@@startlink{#1}\@@href}%
\providecommand \@@href[1]{\endgroup#1\@@endlink}%
\providecommand \@sanitize@url [0]{\catcode `\\12\catcode `\$12\catcode `\&12\catcode `\#12\catcode `\^12\catcode `\_12\catcode `\%12\relax}%
\providecommand \@@startlink[1]{}%
\providecommand \@@endlink[0]{}%
\providecommand \url  [0]{\begingroup\@sanitize@url \@url }%
\providecommand \@url [1]{\endgroup\@href {#1}{\urlprefix }}%
\providecommand \urlprefix  [0]{URL }%
\providecommand \Eprint [0]{\href }%
\providecommand \doibase [0]{https://doi.org/}%
\providecommand \selectlanguage [0]{\@gobble}%
\providecommand \bibinfo  [0]{\@secondoftwo}%
\providecommand \bibfield  [0]{\@secondoftwo}%
\providecommand \translation [1]{[#1]}%
\providecommand \BibitemOpen [0]{}%
\providecommand \bibitemStop [0]{}%
\providecommand \bibitemNoStop [0]{.\EOS\space}%
\providecommand \EOS [0]{\spacefactor3000\relax}%
\providecommand \BibitemShut  [1]{\csname bibitem#1\endcsname}%
\let\auto@bib@innerbib\@empty
\bibitem [{\citenamefont {Montambaux}\ \emph {et~al.}(2009{\natexlab{a}})\citenamefont {Montambaux}, \citenamefont {Pi{\'e}chon}, \citenamefont {Fuchs},\ and\ \citenamefont {Goerbig}}]{montambaux2009universal}%
  \BibitemOpen
  \bibfield  {author} {\bibinfo {author} {\bibfnamefont {G.}~\bibnamefont {Montambaux}}, \bibinfo {author} {\bibfnamefont {F.}~\bibnamefont {Pi{\'e}chon}}, \bibinfo {author} {\bibfnamefont {J.-N.}\ \bibnamefont {Fuchs}},\ and\ \bibinfo {author} {\bibfnamefont {M.}~\bibnamefont {Goerbig}},\ }\bibfield  {title} {\bibinfo {title} {{A universal Hamiltonian for motion and merging of Dirac points in a two-dimensional crystal}},\ }\href {https://link.springer.com/article/10.1140/epjb/e2009-00383-0} {\bibfield  {journal} {\bibinfo  {journal} {The European Physical Journal B}\ }\textbf {\bibinfo {volume} {72}},\ \bibinfo {pages} {509} (\bibinfo {year} {2009}{\natexlab{a}})}\BibitemShut {NoStop}%
\bibitem [{\citenamefont {Bena}\ and\ \citenamefont {Simon}(2011)}]{Bena2011}%
  \BibitemOpen
  \bibfield  {author} {\bibinfo {author} {\bibfnamefont {C.}~\bibnamefont {Bena}}\ and\ \bibinfo {author} {\bibfnamefont {L.}~\bibnamefont {Simon}},\ }\bibfield  {title} {\bibinfo {title} {Dirac point metamorphosis from third-neighbor couplings in graphene and related materials},\ }\href {https://doi.org/10.1103/PhysRevB.83.115404} {\bibfield  {journal} {\bibinfo  {journal} {Phys. Rev. B}\ }\textbf {\bibinfo {volume} {83}},\ \bibinfo {pages} {115404} (\bibinfo {year} {2011})}\BibitemShut {NoStop}%
\bibitem [{\citenamefont {Dietl}\ \emph {et~al.}(2008)\citenamefont {Dietl}, \citenamefont {Pi{\'e}chon},\ and\ \citenamefont {Montambaux}}]{dietl2008new}%
  \BibitemOpen
  \bibfield  {author} {\bibinfo {author} {\bibfnamefont {P.}~\bibnamefont {Dietl}}, \bibinfo {author} {\bibfnamefont {F.}~\bibnamefont {Pi{\'e}chon}},\ and\ \bibinfo {author} {\bibfnamefont {G.}~\bibnamefont {Montambaux}},\ }\bibfield  {title} {\bibinfo {title} {{New magnetic field dependence of Landau levels in a graphenelike structure}},\ }\href {https://journals.aps.org/prl/abstract/10.1103/PhysRevLett.100.236405} {\bibfield  {journal} {\bibinfo  {journal} {Physical review letters}\ }\textbf {\bibinfo {volume} {100}},\ \bibinfo {pages} {236405} (\bibinfo {year} {2008})}\BibitemShut {NoStop}%
\bibitem [{\citenamefont {Montambaux}\ \emph {et~al.}(2009{\natexlab{b}})\citenamefont {Montambaux}, \citenamefont {Pi{\'e}chon}, \citenamefont {Fuchs},\ and\ \citenamefont {Goerbig}}]{montambaux2009merging}%
  \BibitemOpen
  \bibfield  {author} {\bibinfo {author} {\bibfnamefont {G.}~\bibnamefont {Montambaux}}, \bibinfo {author} {\bibfnamefont {F.}~\bibnamefont {Pi{\'e}chon}}, \bibinfo {author} {\bibfnamefont {J.-N.}\ \bibnamefont {Fuchs}},\ and\ \bibinfo {author} {\bibfnamefont {M.~O.}\ \bibnamefont {Goerbig}},\ }\bibfield  {title} {\bibinfo {title} {{Merging of Dirac points in a two-dimensional crystal}},\ }\href {https://journals.aps.org/prb/abstract/10.1103/PhysRevB.80.153412} {\bibfield  {journal} {\bibinfo  {journal} {Physical Review B - Condensed Matter and Materials Physics}\ }\textbf {\bibinfo {volume} {80}},\ \bibinfo {pages} {153412} (\bibinfo {year} {2009}{\natexlab{b}})}\BibitemShut {NoStop}%
\bibitem [{\citenamefont {Kim}\ \emph {et~al.}(2015)\citenamefont {Kim}, \citenamefont {Baik}, \citenamefont {Ryu}, \citenamefont {Sohn}, \citenamefont {Park}, \citenamefont {Park}, \citenamefont {Denlinger}, \citenamefont {Yi}, \citenamefont {Choi},\ and\ \citenamefont {Kim}}]{kim2015observation}%
  \BibitemOpen
  \bibfield  {author} {\bibinfo {author} {\bibfnamefont {J.}~\bibnamefont {Kim}}, \bibinfo {author} {\bibfnamefont {S.~S.}\ \bibnamefont {Baik}}, \bibinfo {author} {\bibfnamefont {S.~H.}\ \bibnamefont {Ryu}}, \bibinfo {author} {\bibfnamefont {Y.}~\bibnamefont {Sohn}}, \bibinfo {author} {\bibfnamefont {S.}~\bibnamefont {Park}}, \bibinfo {author} {\bibfnamefont {B.-G.}\ \bibnamefont {Park}}, \bibinfo {author} {\bibfnamefont {J.}~\bibnamefont {Denlinger}}, \bibinfo {author} {\bibfnamefont {Y.}~\bibnamefont {Yi}}, \bibinfo {author} {\bibfnamefont {H.~J.}\ \bibnamefont {Choi}},\ and\ \bibinfo {author} {\bibfnamefont {K.~S.}\ \bibnamefont {Kim}},\ }\bibfield  {title} {\bibinfo {title} {{Observation of tunable band gap and anisotropic Dirac semimetal state in black phosphorus}},\ }\href {https://www.science.org/doi/full/10.1126/science.aaa6486?casa_token=GI1pQIo97AkAAAAA%3ArTc3phr8WzR7371fIB2FhBKyPMHw1ZOcSCwiPN32iGavCqViq6I1FCqzqVekJ5Cn-kmJ2qfjL8HZ} {\bibfield  {journal} {\bibinfo  {journal} {Science}\ }\textbf
  {\bibinfo {volume} {349}},\ \bibinfo {pages} {723} (\bibinfo {year} {2015})}\BibitemShut {NoStop}%
\bibitem [{\citenamefont {Kim}\ \emph {et~al.}(2017)\citenamefont {Kim}, \citenamefont {Baik}, \citenamefont {Jung}, \citenamefont {Sohn}, \citenamefont {Ryu}, \citenamefont {Choi}, \citenamefont {Yang},\ and\ \citenamefont {Kim}}]{kim2017two}%
  \BibitemOpen
  \bibfield  {author} {\bibinfo {author} {\bibfnamefont {J.}~\bibnamefont {Kim}}, \bibinfo {author} {\bibfnamefont {S.~S.}\ \bibnamefont {Baik}}, \bibinfo {author} {\bibfnamefont {S.~W.}\ \bibnamefont {Jung}}, \bibinfo {author} {\bibfnamefont {Y.}~\bibnamefont {Sohn}}, \bibinfo {author} {\bibfnamefont {S.~H.}\ \bibnamefont {Ryu}}, \bibinfo {author} {\bibfnamefont {H.~J.}\ \bibnamefont {Choi}}, \bibinfo {author} {\bibfnamefont {B.-J.}\ \bibnamefont {Yang}},\ and\ \bibinfo {author} {\bibfnamefont {K.~S.}\ \bibnamefont {Kim}},\ }\bibfield  {title} {\bibinfo {title} {{Two-dimensional dirac fermions protected by space-time inversion symmetry in black phosphorus}},\ }\href {https://journals.aps.org/prl/abstract/10.1103/PhysRevLett.119.226801} {\bibfield  {journal} {\bibinfo  {journal} {Physical review letters}\ }\textbf {\bibinfo {volume} {119}},\ \bibinfo {pages} {226801} (\bibinfo {year} {2017})}\BibitemShut {NoStop}%
\bibitem [{\citenamefont {Shao}\ \emph {et~al.}(2020)\citenamefont {Shao}, \citenamefont {Rudenko}, \citenamefont {Hu}, \citenamefont {Sun}, \citenamefont {Zhu}, \citenamefont {Moon}, \citenamefont {Millis}, \citenamefont {Yuan}, \citenamefont {Lichtenstein}, \citenamefont {Smirnov} \emph {et~al.}}]{shao2020electronic}%
  \BibitemOpen
  \bibfield  {author} {\bibinfo {author} {\bibfnamefont {Y.}~\bibnamefont {Shao}}, \bibinfo {author} {\bibfnamefont {A.}~\bibnamefont {Rudenko}}, \bibinfo {author} {\bibfnamefont {J.}~\bibnamefont {Hu}}, \bibinfo {author} {\bibfnamefont {Z.}~\bibnamefont {Sun}}, \bibinfo {author} {\bibfnamefont {Y.}~\bibnamefont {Zhu}}, \bibinfo {author} {\bibfnamefont {S.}~\bibnamefont {Moon}}, \bibinfo {author} {\bibfnamefont {A.}~\bibnamefont {Millis}}, \bibinfo {author} {\bibfnamefont {S.}~\bibnamefont {Yuan}}, \bibinfo {author} {\bibfnamefont {A.}~\bibnamefont {Lichtenstein}}, \bibinfo {author} {\bibfnamefont {D.}~\bibnamefont {Smirnov}}, \emph {et~al.},\ }\bibfield  {title} {\bibinfo {title} {{Electronic correlations in nodal-line semimetals}},\ }\href {https://www.nature.com/articles/s41567-020-0859-z} {\bibfield  {journal} {\bibinfo  {journal} {Nature Physics}\ }\textbf {\bibinfo {volume} {16}},\ \bibinfo {pages} {636} (\bibinfo {year} {2020})}\BibitemShut {NoStop}%
\bibitem [{\citenamefont {Shao}\ \emph {et~al.}(2024)\citenamefont {Shao}, \citenamefont {Moon}, \citenamefont {Rudenko}, \citenamefont {Wang}, \citenamefont {Herzog-Arbeitman}, \citenamefont {Ozerov}, \citenamefont {Graf}, \citenamefont {Sun}, \citenamefont {Queiroz}, \citenamefont {Lee}, \citenamefont {Zhu}, \citenamefont {Mao}, \citenamefont {Katsnelson}, \citenamefont {Bernevig}, \citenamefont {Smirnov}, \citenamefont {Millis},\ and\ \citenamefont {Basov}}]{shao2024}%
  \BibitemOpen
  \bibfield  {author} {\bibinfo {author} {\bibfnamefont {Y.}~\bibnamefont {Shao}}, \bibinfo {author} {\bibfnamefont {S.}~\bibnamefont {Moon}}, \bibinfo {author} {\bibfnamefont {A.~N.}\ \bibnamefont {Rudenko}}, \bibinfo {author} {\bibfnamefont {J.}~\bibnamefont {Wang}}, \bibinfo {author} {\bibfnamefont {J.}~\bibnamefont {Herzog-Arbeitman}}, \bibinfo {author} {\bibfnamefont {M.}~\bibnamefont {Ozerov}}, \bibinfo {author} {\bibfnamefont {D.}~\bibnamefont {Graf}}, \bibinfo {author} {\bibfnamefont {Z.}~\bibnamefont {Sun}}, \bibinfo {author} {\bibfnamefont {R.}~\bibnamefont {Queiroz}}, \bibinfo {author} {\bibfnamefont {S.~H.}\ \bibnamefont {Lee}}, \bibinfo {author} {\bibfnamefont {Y.}~\bibnamefont {Zhu}}, \bibinfo {author} {\bibfnamefont {Z.}~\bibnamefont {Mao}}, \bibinfo {author} {\bibfnamefont {M.~I.}\ \bibnamefont {Katsnelson}}, \bibinfo {author} {\bibfnamefont {B.~A.}\ \bibnamefont {Bernevig}}, \bibinfo {author} {\bibfnamefont {D.}~\bibnamefont {Smirnov}}, \bibinfo {author} {\bibfnamefont {A.~J.}\ \bibnamefont
  {Millis}},\ and\ \bibinfo {author} {\bibfnamefont {D.~N.}\ \bibnamefont {Basov}},\ }\bibfield  {title} {\bibinfo {title} {Semi-dirac fermions in a topological metal},\ }\href {https://doi.org/10.1103/PhysRevX.14.041057} {\bibfield  {journal} {\bibinfo  {journal} {Phys. Rev. X}\ }\textbf {\bibinfo {volume} {14}},\ \bibinfo {pages} {041057} (\bibinfo {year} {2024})}\BibitemShut {NoStop}%
\bibitem [{\citenamefont {Amorim}\ \emph {et~al.}(2016)\citenamefont {Amorim}, \citenamefont {Cortijo}, \citenamefont {De~Juan}, \citenamefont {Grushin}, \citenamefont {Guinea}, \citenamefont {Guti{\'e}rrez-Rubio}, \citenamefont {Ochoa}, \citenamefont {Parente}, \citenamefont {Rold{\'a}n}, \citenamefont {San-Jose} \emph {et~al.}}]{amorim2016novel}%
  \BibitemOpen
  \bibfield  {author} {\bibinfo {author} {\bibfnamefont {B.}~\bibnamefont {Amorim}}, \bibinfo {author} {\bibfnamefont {A.}~\bibnamefont {Cortijo}}, \bibinfo {author} {\bibfnamefont {F.}~\bibnamefont {De~Juan}}, \bibinfo {author} {\bibfnamefont {A.~G.}\ \bibnamefont {Grushin}}, \bibinfo {author} {\bibfnamefont {F.}~\bibnamefont {Guinea}}, \bibinfo {author} {\bibfnamefont {A.}~\bibnamefont {Guti{\'e}rrez-Rubio}}, \bibinfo {author} {\bibfnamefont {H.}~\bibnamefont {Ochoa}}, \bibinfo {author} {\bibfnamefont {V.}~\bibnamefont {Parente}}, \bibinfo {author} {\bibfnamefont {R.}~\bibnamefont {Rold{\'a}n}}, \bibinfo {author} {\bibfnamefont {P.}~\bibnamefont {San-Jose}}, \emph {et~al.},\ }\bibfield  {title} {\bibinfo {title} {{Novel effects of strains in graphene and other two dimensional materials}},\ }\href {https://www.sciencedirect.com/science/article/pii/S0370157315005402?casa_token=wuNVzn3VrAkAAAAA:DswyTE7d7CXVEIscsriucCniWcwjWLb-bTlrag6f7GV4gfEqqU4rNb7bhzYJ_KuGhVsyp1Mt} {\bibfield  {journal} {\bibinfo  {journal}
  {Physics Reports}\ }\textbf {\bibinfo {volume} {617}},\ \bibinfo {pages} {1} (\bibinfo {year} {2016})}\BibitemShut {NoStop}%
\bibitem [{\citenamefont {Katayama}\ \emph {et~al.}(2006)\citenamefont {Katayama}, \citenamefont {Kobayashi},\ and\ \citenamefont {Suzumura}}]{katayama2006pressure}%
  \BibitemOpen
  \bibfield  {author} {\bibinfo {author} {\bibfnamefont {S.}~\bibnamefont {Katayama}}, \bibinfo {author} {\bibfnamefont {A.}~\bibnamefont {Kobayashi}},\ and\ \bibinfo {author} {\bibfnamefont {Y.}~\bibnamefont {Suzumura}},\ }\bibfield  {title} {\bibinfo {title} {{Pressure-induced zero-gap semiconducting state in organic conductor $\alpha$-(BEDT-TTF) 2I3 salt}},\ }\href {https://journals.jps.jp/doi/abs/10.1143/JPSJ.75.054705} {\bibfield  {journal} {\bibinfo  {journal} {Journal of the Physical Society of Japan}\ }\textbf {\bibinfo {volume} {75}},\ \bibinfo {pages} {054705} (\bibinfo {year} {2006})}\BibitemShut {NoStop}%
\bibitem [{\citenamefont {Subedi}\ and\ \citenamefont {Behnia}(2024)}]{subedi2024semidiracfermionshexagonalclosepacked}%
  \BibitemOpen
  \bibfield  {author} {\bibinfo {author} {\bibfnamefont {A.}~\bibnamefont {Subedi}}\ and\ \bibinfo {author} {\bibfnamefont {K.}~\bibnamefont {Behnia}},\ }\href {https://arxiv.org/abs/2411.11585} {\bibinfo {title} {{Semi-Dirac fermions in hexagonal close-packed cadmium}}} (\bibinfo {year} {2024}),\ \Eprint {https://arxiv.org/abs/2411.11585} {arXiv:2411.11585 [cond-mat.mtrl-sci]} \BibitemShut {NoStop}%
\bibitem [{\citenamefont {Pardo}\ and\ \citenamefont {Pickett}(2009)}]{pardo2009half}%
  \BibitemOpen
  \bibfield  {author} {\bibinfo {author} {\bibfnamefont {V.}~\bibnamefont {Pardo}}\ and\ \bibinfo {author} {\bibfnamefont {W.~E.}\ \bibnamefont {Pickett}},\ }\bibfield  {title} {\bibinfo {title} {{Half-metallic semi-Dirac-point generated by quantum confinement in TiO$_2$/VO$_2$ nanostructures}},\ }\href {https://journals.aps.org/prl/abstract/10.1103/PhysRevLett.102.166803} {\bibfield  {journal} {\bibinfo  {journal} {Physical review letters}\ }\textbf {\bibinfo {volume} {102}},\ \bibinfo {pages} {166803} (\bibinfo {year} {2009})}\BibitemShut {NoStop}%
\bibitem [{\citenamefont {Huang}\ \emph {et~al.}(2015)\citenamefont {Huang}, \citenamefont {Liu}, \citenamefont {Zhang}, \citenamefont {Duan},\ and\ \citenamefont {Vanderbilt}}]{huang2015emergence}%
  \BibitemOpen
  \bibfield  {author} {\bibinfo {author} {\bibfnamefont {H.}~\bibnamefont {Huang}}, \bibinfo {author} {\bibfnamefont {Z.}~\bibnamefont {Liu}}, \bibinfo {author} {\bibfnamefont {H.}~\bibnamefont {Zhang}}, \bibinfo {author} {\bibfnamefont {W.}~\bibnamefont {Duan}},\ and\ \bibinfo {author} {\bibfnamefont {D.}~\bibnamefont {Vanderbilt}},\ }\bibfield  {title} {\bibinfo {title} {{Emergence of a Chern-insulating state from a semi-Dirac dispersion}},\ }\href {https://journals.aps.org/prb/abstract/10.1103/PhysRevB.92.161115} {\bibfield  {journal} {\bibinfo  {journal} {Physical Review B}\ }\textbf {\bibinfo {volume} {92}},\ \bibinfo {pages} {161115} (\bibinfo {year} {2015})}\BibitemShut {NoStop}%
\bibitem [{\citenamefont {Uchoa}\ and\ \citenamefont {Seo}(2017)}]{uchoa2017superconducting}%
  \BibitemOpen
  \bibfield  {author} {\bibinfo {author} {\bibfnamefont {B.}~\bibnamefont {Uchoa}}\ and\ \bibinfo {author} {\bibfnamefont {K.}~\bibnamefont {Seo}},\ }\bibfield  {title} {\bibinfo {title} {{Superconducting states for semi-Dirac fermions at zero and finite magnetic fields}},\ }\href {https://journals.aps.org/prb/abstract/10.1103/PhysRevB.96.220503} {\bibfield  {journal} {\bibinfo  {journal} {Physical Review B}\ }\textbf {\bibinfo {volume} {96}},\ \bibinfo {pages} {220503} (\bibinfo {year} {2017})}\BibitemShut {NoStop}%
\bibitem [{\citenamefont {Uryszek}\ \emph {et~al.}(2019)\citenamefont {Uryszek}, \citenamefont {Christou}, \citenamefont {Jaefari}, \citenamefont {Kr{\"u}ger},\ and\ \citenamefont {Uchoa}}]{uryszek2019quantum}%
  \BibitemOpen
  \bibfield  {author} {\bibinfo {author} {\bibfnamefont {M.~D.}\ \bibnamefont {Uryszek}}, \bibinfo {author} {\bibfnamefont {E.}~\bibnamefont {Christou}}, \bibinfo {author} {\bibfnamefont {A.}~\bibnamefont {Jaefari}}, \bibinfo {author} {\bibfnamefont {F.}~\bibnamefont {Kr{\"u}ger}},\ and\ \bibinfo {author} {\bibfnamefont {B.}~\bibnamefont {Uchoa}},\ }\bibfield  {title} {\bibinfo {title} {{Quantum criticality of semi-Dirac fermions in 2+ 1 dimensions}},\ }\href {https://journals.aps.org/prb/abstract/10.1103/PhysRevB.100.155101} {\bibfield  {journal} {\bibinfo  {journal} {Physical Review B}\ }\textbf {\bibinfo {volume} {100}},\ \bibinfo {pages} {155101} (\bibinfo {year} {2019})}\BibitemShut {NoStop}%
\bibitem [{\citenamefont {Uryszek}\ \emph {et~al.}(2020)\citenamefont {Uryszek}, \citenamefont {Kr{\"u}ger},\ and\ \citenamefont {Christou}}]{uryszek2020fermionic}%
  \BibitemOpen
  \bibfield  {author} {\bibinfo {author} {\bibfnamefont {M.~D.}\ \bibnamefont {Uryszek}}, \bibinfo {author} {\bibfnamefont {F.}~\bibnamefont {Kr{\"u}ger}},\ and\ \bibinfo {author} {\bibfnamefont {E.}~\bibnamefont {Christou}},\ }\bibfield  {title} {\bibinfo {title} {{Fermionic criticality of anisotropic nodal point semimetals away from the upper critical dimension: Exact exponents to leading order in $1/ N_f$}},\ }\href {https://journals.aps.org/prresearch/abstract/10.1103/PhysRevResearch.2.043265} {\bibfield  {journal} {\bibinfo  {journal} {Physical Review Research}\ }\textbf {\bibinfo {volume} {2}},\ \bibinfo {pages} {043265} (\bibinfo {year} {2020})}\BibitemShut {NoStop}%
\bibitem [{\citenamefont {Roy}\ and\ \citenamefont {Foster}(2018)}]{roy2018quantum}%
  \BibitemOpen
  \bibfield  {author} {\bibinfo {author} {\bibfnamefont {B.}~\bibnamefont {Roy}}\ and\ \bibinfo {author} {\bibfnamefont {M.~S.}\ \bibnamefont {Foster}},\ }\bibfield  {title} {\bibinfo {title} {{Quantum multicriticality near the Dirac-semimetal to band-insulator critical point in two dimensions: A controlled ascent from one dimension}},\ }\href {https://journals.aps.org/prx/abstract/10.1103/PhysRevX.8.011049} {\bibfield  {journal} {\bibinfo  {journal} {Physical Review X}\ }\textbf {\bibinfo {volume} {8}},\ \bibinfo {pages} {011049} (\bibinfo {year} {2018})}\BibitemShut {NoStop}%
\bibitem [{\citenamefont {Kotov}\ \emph {et~al.}(2012)\citenamefont {Kotov}, \citenamefont {Uchoa}, \citenamefont {Pereira}, \citenamefont {Guinea},\ and\ \citenamefont {Castro~Neto}}]{kotov2012electron}%
  \BibitemOpen
  \bibfield  {author} {\bibinfo {author} {\bibfnamefont {V.~N.}\ \bibnamefont {Kotov}}, \bibinfo {author} {\bibfnamefont {B.}~\bibnamefont {Uchoa}}, \bibinfo {author} {\bibfnamefont {V.~M.}\ \bibnamefont {Pereira}}, \bibinfo {author} {\bibfnamefont {F.}~\bibnamefont {Guinea}},\ and\ \bibinfo {author} {\bibfnamefont {A.}~\bibnamefont {Castro~Neto}},\ }\bibfield  {title} {\bibinfo {title} {{Electron-electron interactions in graphene: Current status and perspectives}},\ }\href {https://journals.aps.org/rmp/abstract/10.1103/RevModPhys.84.1067} {\bibfield  {journal} {\bibinfo  {journal} {Reviews of modern physics}\ }\textbf {\bibinfo {volume} {84}},\ \bibinfo {pages} {1067} (\bibinfo {year} {2012})}\BibitemShut {NoStop}%
\bibitem [{\citenamefont {Isobe}\ \emph {et~al.}(2016)\citenamefont {Isobe}, \citenamefont {Yang}, \citenamefont {Chubukov}, \citenamefont {Schmalian},\ and\ \citenamefont {Nagaosa}}]{isobe2016emergent}%
  \BibitemOpen
  \bibfield  {author} {\bibinfo {author} {\bibfnamefont {H.}~\bibnamefont {Isobe}}, \bibinfo {author} {\bibfnamefont {B.-J.}\ \bibnamefont {Yang}}, \bibinfo {author} {\bibfnamefont {A.}~\bibnamefont {Chubukov}}, \bibinfo {author} {\bibfnamefont {J.}~\bibnamefont {Schmalian}},\ and\ \bibinfo {author} {\bibfnamefont {N.}~\bibnamefont {Nagaosa}},\ }\bibfield  {title} {\bibinfo {title} {{Emergent non-fermi-liquid at the quantum critical point of a topological phase transition in two dimensions}},\ }\href {https://journals.aps.org/prl/abstract/10.1103/PhysRevLett.116.076803} {\bibfield  {journal} {\bibinfo  {journal} {Physical review letters}\ }\textbf {\bibinfo {volume} {116}},\ \bibinfo {pages} {076803} (\bibinfo {year} {2016})}\BibitemShut {NoStop}%
\bibitem [{\citenamefont {Cho}\ and\ \citenamefont {Moon}(2016)}]{cho2016novel}%
  \BibitemOpen
  \bibfield  {author} {\bibinfo {author} {\bibfnamefont {G.~Y.}\ \bibnamefont {Cho}}\ and\ \bibinfo {author} {\bibfnamefont {E.-G.}\ \bibnamefont {Moon}},\ }\bibfield  {title} {\bibinfo {title} {{Novel quantum criticality in two dimensional topological phase transitions}},\ }\href {https://www.nature.com/articles/srep19198} {\bibfield  {journal} {\bibinfo  {journal} {Scientific Reports}\ }\textbf {\bibinfo {volume} {6}},\ \bibinfo {pages} {19198} (\bibinfo {year} {2016})}\BibitemShut {NoStop}%
\bibitem [{\citenamefont {Kotov}\ \emph {et~al.}(2021)\citenamefont {Kotov}, \citenamefont {Uchoa},\ and\ \citenamefont {Sushkov}}]{kotov2021coulomb}%
  \BibitemOpen
  \bibfield  {author} {\bibinfo {author} {\bibfnamefont {V.~N.}\ \bibnamefont {Kotov}}, \bibinfo {author} {\bibfnamefont {B.}~\bibnamefont {Uchoa}},\ and\ \bibinfo {author} {\bibfnamefont {O.~P.}\ \bibnamefont {Sushkov}},\ }\bibfield  {title} {\bibinfo {title} {{Coulomb interactions and renormalization of semi-Dirac fermions near a topological Lifshitz transition}},\ }\href {https://journals.aps.org/prb/abstract/10.1103/PhysRevB.103.045403} {\bibfield  {journal} {\bibinfo  {journal} {Physical Review B}\ }\textbf {\bibinfo {volume} {103}},\ \bibinfo {pages} {045403} (\bibinfo {year} {2021})}\BibitemShut {NoStop}%
\bibitem [{\citenamefont {Quan}\ and\ \citenamefont {Pickett}(2018)}]{quan2018maximally}%
  \BibitemOpen
  \bibfield  {author} {\bibinfo {author} {\bibfnamefont {Y.}~\bibnamefont {Quan}}\ and\ \bibinfo {author} {\bibfnamefont {W.~E.}\ \bibnamefont {Pickett}},\ }\bibfield  {title} {\bibinfo {title} {{A maximally particle-hole asymmetric spectrum emanating from a semi-Dirac point}},\ }\href {https://iopscience.iop.org/article/10.1088/1361-648X/aaa521/meta} {\bibfield  {journal} {\bibinfo  {journal} {Journal of Physics: Condensed Matter}\ }\textbf {\bibinfo {volume} {30}},\ \bibinfo {pages} {075501} (\bibinfo {year} {2018})}\BibitemShut {NoStop}%
\bibitem [{\citenamefont {Carbotte}\ \emph {et~al.}(2019)\citenamefont {Carbotte}, \citenamefont {Bryenton},\ and\ \citenamefont {Nicol}}]{carbotte2019optical}%
  \BibitemOpen
  \bibfield  {author} {\bibinfo {author} {\bibfnamefont {J.}~\bibnamefont {Carbotte}}, \bibinfo {author} {\bibfnamefont {K.}~\bibnamefont {Bryenton}},\ and\ \bibinfo {author} {\bibfnamefont {E.}~\bibnamefont {Nicol}},\ }\bibfield  {title} {\bibinfo {title} {{Optical properties of a semi-Dirac material}},\ }\href {https://journals.aps.org/prb/abstract/10.1103/PhysRevB.99.115406} {\bibfield  {journal} {\bibinfo  {journal} {Physical Review B}\ }\textbf {\bibinfo {volume} {99}},\ \bibinfo {pages} {115406} (\bibinfo {year} {2019})}\BibitemShut {NoStop}%
\bibitem [{\citenamefont {Foster}\ and\ \citenamefont {Aleiner}(2008)}]{foster2008graphene}%
  \BibitemOpen
  \bibfield  {author} {\bibinfo {author} {\bibfnamefont {M.~S.}\ \bibnamefont {Foster}}\ and\ \bibinfo {author} {\bibfnamefont {I.~L.}\ \bibnamefont {Aleiner}},\ }\bibfield  {title} {\bibinfo {title} {{Graphene via large N: A renormalization group study}},\ }\href {https://journals.aps.org/prb/abstract/10.1103/PhysRevB.77.195413} {\bibfield  {journal} {\bibinfo  {journal} {Physical Review B—Condensed Matter and Materials Physics}\ }\textbf {\bibinfo {volume} {77}},\ \bibinfo {pages} {195413} (\bibinfo {year} {2008})}\BibitemShut {NoStop}%
\bibitem [{\citenamefont {Nandkishore}\ and\ \citenamefont {Levitov}(2010)}]{nandkishore2010electron}%
  \BibitemOpen
  \bibfield  {author} {\bibinfo {author} {\bibfnamefont {R.}~\bibnamefont {Nandkishore}}\ and\ \bibinfo {author} {\bibfnamefont {L.}~\bibnamefont {Levitov}},\ }\bibfield  {title} {\bibinfo {title} {{Electron interactions in bilayer graphene: Marginal Fermi liquid and zero-bias anomaly}},\ }\href {https://journals.aps.org/prb/abstract/10.1103/PhysRevB.82.115431} {\bibfield  {journal} {\bibinfo  {journal} {Physical Review B—Condensed Matter and Materials Physics}\ }\textbf {\bibinfo {volume} {82}},\ \bibinfo {pages} {115431} (\bibinfo {year} {2010})}\BibitemShut {NoStop}%
\bibitem [{\citenamefont {Dou}\ \emph {et~al.}(2014)\citenamefont {Dou}, \citenamefont {Jaefari}, \citenamefont {Barlas},\ and\ \citenamefont {Uchoa}}]{dou2014quasiparticle}%
  \BibitemOpen
  \bibfield  {author} {\bibinfo {author} {\bibfnamefont {X.}~\bibnamefont {Dou}}, \bibinfo {author} {\bibfnamefont {A.}~\bibnamefont {Jaefari}}, \bibinfo {author} {\bibfnamefont {Y.}~\bibnamefont {Barlas}},\ and\ \bibinfo {author} {\bibfnamefont {B.}~\bibnamefont {Uchoa}},\ }\bibfield  {title} {\bibinfo {title} {{Quasiparticle renormalization in ABC graphene trilayers}},\ }\href {https://journals.aps.org/prb/abstract/10.1103/PhysRevB.90.161411} {\bibfield  {journal} {\bibinfo  {journal} {Physical Review B}\ }\textbf {\bibinfo {volume} {90}},\ \bibinfo {pages} {161411} (\bibinfo {year} {2014})}\BibitemShut {NoStop}%
\bibitem [{\citenamefont {Bostwick}\ \emph {et~al.}(2007)\citenamefont {Bostwick}, \citenamefont {Ohta}, \citenamefont {Seyller}, \citenamefont {Horn},\ and\ \citenamefont {Rotenberg}}]{bostwick2007quasiparticle}%
  \BibitemOpen
  \bibfield  {author} {\bibinfo {author} {\bibfnamefont {A.}~\bibnamefont {Bostwick}}, \bibinfo {author} {\bibfnamefont {T.}~\bibnamefont {Ohta}}, \bibinfo {author} {\bibfnamefont {T.}~\bibnamefont {Seyller}}, \bibinfo {author} {\bibfnamefont {K.}~\bibnamefont {Horn}},\ and\ \bibinfo {author} {\bibfnamefont {E.}~\bibnamefont {Rotenberg}},\ }\bibfield  {title} {\bibinfo {title} {{Quasiparticle dynamics in graphene}},\ }\href {https://www.nature.com/articles/nphys477} {\bibfield  {journal} {\bibinfo  {journal} {Nature physics}\ }\textbf {\bibinfo {volume} {3}},\ \bibinfo {pages} {36} (\bibinfo {year} {2007})}\BibitemShut {NoStop}%
\bibitem [{\citenamefont {Martin}\ \emph {et~al.}(2008)\citenamefont {Martin}, \citenamefont {Akerman}, \citenamefont {Ulbricht}, \citenamefont {Lohmann}, \citenamefont {Smet}, \citenamefont {Von~Klitzing},\ and\ \citenamefont {Yacoby}}]{martin2008observation}%
  \BibitemOpen
  \bibfield  {author} {\bibinfo {author} {\bibfnamefont {J.}~\bibnamefont {Martin}}, \bibinfo {author} {\bibfnamefont {N.}~\bibnamefont {Akerman}}, \bibinfo {author} {\bibfnamefont {G.}~\bibnamefont {Ulbricht}}, \bibinfo {author} {\bibfnamefont {T.}~\bibnamefont {Lohmann}}, \bibinfo {author} {\bibfnamefont {J.~v.}\ \bibnamefont {Smet}}, \bibinfo {author} {\bibfnamefont {K.}~\bibnamefont {Von~Klitzing}},\ and\ \bibinfo {author} {\bibfnamefont {A.}~\bibnamefont {Yacoby}},\ }\bibfield  {title} {\bibinfo {title} {{Observation of electron--hole puddles in graphene using a scanning single-electron transistor}},\ }\href {https://www.nature.com/articles/nphys781} {\bibfield  {journal} {\bibinfo  {journal} {Nature physics}\ }\textbf {\bibinfo {volume} {4}},\ \bibinfo {pages} {144} (\bibinfo {year} {2008})}\BibitemShut {NoStop}%
\bibitem [{\citenamefont {Yu}\ \emph {et~al.}(2013)\citenamefont {Yu}, \citenamefont {Jalil}, \citenamefont {Belle}, \citenamefont {Mayorov}, \citenamefont {Blake}, \citenamefont {Schedin}, \citenamefont {Morozov}, \citenamefont {Ponomarenko}, \citenamefont {Chiappini}, \citenamefont {Wiedmann} \emph {et~al.}}]{yu2013interaction}%
  \BibitemOpen
  \bibfield  {author} {\bibinfo {author} {\bibfnamefont {G.}~\bibnamefont {Yu}}, \bibinfo {author} {\bibfnamefont {R.}~\bibnamefont {Jalil}}, \bibinfo {author} {\bibfnamefont {B.}~\bibnamefont {Belle}}, \bibinfo {author} {\bibfnamefont {A.~S.}\ \bibnamefont {Mayorov}}, \bibinfo {author} {\bibfnamefont {P.}~\bibnamefont {Blake}}, \bibinfo {author} {\bibfnamefont {F.}~\bibnamefont {Schedin}}, \bibinfo {author} {\bibfnamefont {S.~V.}\ \bibnamefont {Morozov}}, \bibinfo {author} {\bibfnamefont {L.~A.}\ \bibnamefont {Ponomarenko}}, \bibinfo {author} {\bibfnamefont {F.}~\bibnamefont {Chiappini}}, \bibinfo {author} {\bibfnamefont {S.}~\bibnamefont {Wiedmann}}, \emph {et~al.},\ }\bibfield  {title} {\bibinfo {title} {{Interaction phenomena in graphene seen through quantum capacitance}},\ }\href {https://www.pnas.org/doi/abs/10.1073/pnas.1300599110} {\bibfield  {journal} {\bibinfo  {journal} {Proceedings of the National Academy of Sciences}\ }\textbf {\bibinfo {volume} {110}},\ \bibinfo {pages} {3282} (\bibinfo {year}
  {2013})}\BibitemShut {NoStop}%
\end{thebibliography}%

\end{document}